\documentclass[aps, twocolumn, amsmath, amssymb, prb, eqsecnum, 10pt, floatfix]{revtex4-1}

\usepackage{amsmath,color,colordvi}
\usepackage{mathtools}
\usepackage[pdftex]{graphicx}
\usepackage[english]{babel}
\usepackage{braket}

\definecolor{navyblue}{rgb}{0.0, 0.0, 0.5}
\usepackage[protrusion=true, expansion=true]{microtype} 
\usepackage[colorlinks=true, breaklinks=false, linkcolor=navyblue, urlcolor=navyblue, citecolor=navyblue]{hyperref}

\newcommand{\D}[1]{\text{d}#1}
\newcommand{\ii}[0]{\mathrm{i}}
\newcommand{\ee}[0]{\mathrm{e}}

\newcommand{\updownarrows}{\mathbin\uparrow,\hspace{-.0em}\downarrow}

\newcommand{\normrateright}[0]{\overrightarrow{F}}
\newcommand{\normrateleft}[0]{\overleftarrow{F}}
\newcommand{\nmd}[0]{\rm N}

\graphicspath{{../figures/}}

\begin{document}

\title{Theory of quantum-circuit refrigeration by photon-assisted electron tunneling}

\author{Matti Silveri$^{1,2}$}
\email{matti.silveri@oulu.fi}
\author{Hermann Grabert$^3$}
\author{Shumpei Masuda$^1$}
\author{Kuan Yen Tan$^1$}
\author{Mikko M\"ott\"onen$^{1,4}$}
\email{mikko.mottonen@aalto.fi}

\affiliation{$^1$QCD Labs, COMP Centre of Excellence, Department of Applied Physics, Aalto University, P.O.~Box 13500, FI-00076 Aalto, Finland\\$^2$Research Unit of Nano and Molecular Systems, University of Oulu, P.O. Box 3000, FI-90014 Oulu, Finland\\$^3$Department of Physics, University of Freiburg, D-79104 Freiburg, Germany\\$^4$University of Jyv\"askyl\"a, Department of Mathematical Information Technology, P.O.~Box 35, FI-40014 University of Jyv\"askyl\"a, Finland}

\date{\today}
\selectlanguage{english}

\begin{abstract}
We focus on a recently experimentally realized scenario of normal-metal--insulator--superconductor tunnel junctions coupled to a superconducting resonator. We develop a first-principles theory to describe the effect of photon-assisted electron tunneling on the quantum state of the resonator. Our results are in very good quantitative agreement with the previous experiments on refrigeration and heating of the resonator using the photon-assisted tunneling, thus providing a stringent verification of the developed theory. Importantly, our results provide simple analytical estimates of the voltage-tunable coupling strength and temperature of the thermal reservoir formed by the photon-assisted tunneling.  Consequently, they are used to introduce optimization principles for initialization of quantum devices using such a quantum-circuit refrigerator. Thanks to the first-principles nature of our approach, extension of the theory to the full spectrum of quantum electric devices seems~plausible.
\end{abstract}
\maketitle

\section{Introduction}
Superconducting quantum circuits~\cite{Nakamura99,Vion02,Martinis02,Wallraff04,Manucharyan09,Riste12,Devoret13, Barends2013, Ofek16} are among the leading candidates of quantum technological devices for the implementation of large-scale quantum computing~\cite{Chow14, Kelly15,Versluis16} and simulations~\cite{Houck12}, with envisioned applications of great practical value. However, fast and accurate initialization of these devices to a pure quantum state remains challenging although it is a key requirement in their efficient operation~\cite{Divincenzo00}. A solution could be an active refrigerator~\cite{Grajcar08} which evacuates entropy on demand for efficient initialization. Such device may also provide a route to robust ground-state operation by reduction of errors related to thermal and non-adiabatic excitations.

 \begin{figure}
   \centering
  \includegraphics[width=0.95\linewidth]{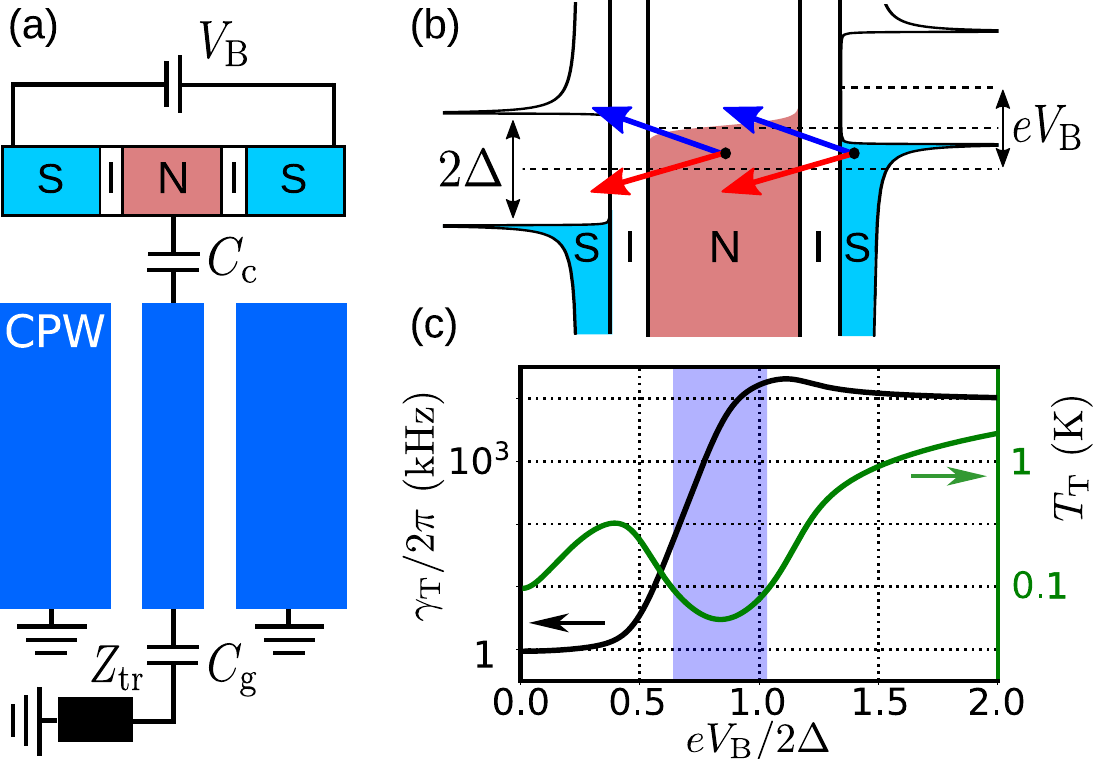}
   \caption{\label{fig:fig1} (a) Schematic diagram of a superconducting coplanar waveguide (CPW) resonator which is connected through the capacitances~$C_{\rm c}$ and~$C_{\rm g}$ to a normal-metal island and a transmission line with an impedance~$Z_{\rm tr}$. The superconductor~(S)--insulator~(I)--normal-metal~(N) tunnel junctions defining the island are voltage biased through the superconducting electrodes. (b)~Energy diagram of the photon-assisted electron tunneling. The blue arrows depict tunneling events leading to absorption of a photon from the coupled resonator and the red arrows correspond to emission.  (c) Coupling strength~$\gamma_{\rm T}$ and temperature~$T_{\rm T}$ of the effective thermal reservoir formed by the photon-assisted tunneling as a function of the two-junction voltage bias. The parameters correspond to typical experimental values: see Fig.~\ref{fig:gammaN}. In the highlighted region the thermal reservoir is cooler than the electrons of the normal-metal island.}
\end{figure}

Much effort has been put into studies of incoherent tunneling of single charges in mesoscopic junctions~\cite{Averin86,Fulton87,Ingold91, Devoret90, Girvin90, IngoldNazarov05, Ingold94, Lee96, Likharev99, Catelani11, Pekola13}. Whereas fully normal-metal junctions can be used as sensitive charge sensors~\cite{Devoret92} and primary thermometers~\cite{Pekola94}, normal-metal--insulator--superconductor (NIS) junctions have opened an avenue for electrically refrigerating the normal-metal electron reservoirs even below the phonon bath temperature~\cite{Giazotto06}. However, quantum devices are designed to be very well isolated from dissipative electron systems owing to the requirements of long coherence time~\cite{Ofek16}, and hence the benefits of the NIS junction technology in quantum-circuit initialization~\cite{Jones13,Tuorila16} are far from obvious.

Although the early work on photon-assisted tunneling at NIS junctions focused on the effect of the electromagnetic circuit on the tunnel current~\cite{Devoret90,Girvin90,IngoldNazarov05,Averin90,Pekola10}, recent studies also demonstrate the effect of the tunneling events on the state of the circuit~\cite{Zakka10,Hofheinz11, Altimiras14, Stockklauser15, Kubala15, Bruhat16, Tan16, Masuda16, Westig17, Jebari17}. Importantly, a quantum-circuit refrigerator and cryogenic microwave source based on photon-assisted tunneling of electrons through NIS junctions were demonstrated in Refs.~\onlinecite{Tan16, Masuda16}, see schematic in Fig.~\ref{fig:fig1}. The refrigeration occurs at junction bias voltages where the normal-metal electron needs to receive an additional energy quantum from the coupled quantum electric circuit to overcome the Bardeen--Cooper--Schrieffer energy gap in the superconductor, see Fig.~\ref{fig:fig1}(b)-(c). The resulting exponential tunability of the coupling strength with the bias voltage offers a promising technique to quickly initialize quantum systems on demand. Furthermore, when the junctions are biased above the superconductor gap, tunneling events which emit additional energy to the coupled quantum circuit become energetically allowed and the device can be utilized as a source of incoherent microwave radiation~\cite{Masuda16}.

In this paper, we provide a first-principles derivation of the relaxation and excitation rates induced on a superconducting resonator which is capacitively coupled to NIS junctions. This model accurately describes the physics of the quantum-circuit refrigerator and the cryogenic microwave source demonstrated in Refs.~\onlinecite{Tan16, Masuda16}. In contrast to the previous model~\cite{Tan16, Masuda16}, we are able to capture fine details of the physical circuit, multi-photon states, and multi-photon absorption. Importantly, we put our model to an experimental test by directly comparing the recently measured radiation generated by the NIS junctions at high bias voltages~\cite{Masuda16} to that predicted by our model. The obtained excellent agreement verifies that our approach is valid, and encourages further extension of the theory to the full spectrum of quantum devices.

This paper is organized as follows: Section~\ref{sec:ExpScenario} presents the physical system under study and Sec.~\ref{sec:ModHamil} introduces the corresponding Hamiltonian operators.  In Sec.~\ref{sec:Tunnel}, we diagonalize the system Hamiltonian and derive the tunneling-induced transition rates between the eigenstates. Section~\ref{sec:Master} provides a master equation for the resonator and a thermal reservoir model of the photon-assisted tunneling. Section~\ref{sec:ExpLimits} is devoted to analytical approximations of the coupling strength and temperature of the thermal reservoir at different bias voltage regimes. In Sec.~\ref{sec:QCRcomp}, we present optimal parameters for using the quantum-circuit refrigerator for cooling.  In Sec.~\ref{sec:heatingcomp}, we study the heating regime at high bias voltages and compare our results with the measurements of Ref.~\onlinecite{Masuda16} and with the previous theoretical model based on $P(E)$-theory~\cite{Devoret90, Girvin90, IngoldNazarov05}. Section~\ref{sec:conc} provides  our conclusions and an outlook into the future of quantum-circuit refrigeration.

\section{Experimental scenario}\label{sec:ExpScenario}
The physical system studied in this paper is illustrated in Fig.~\ref{fig:fig1}(a). The central element is a coplanar waveguide resonator with the fundamental resonance angular frequency~$\omega_{\rm r}$. At one end, the resonator is capacitively coupled to a normal-metal island which is equipped with two identical normal-metal--insulator--superconductor junctions. At the other end, the resonator is capacitively coupled to a transmission line of characteristic impedance~$Z_{\rm tr}$.

A pair of NIS junctions, \textit{i.e.}, a superconductor--insulator--normal-metal--insulator--superconductor (SINIS) junction, is biased by a voltage~$V_{\rm B}=2V$, where $V$ is the bias of a single NIS junction. This allows for a voltage-controlled charging and discharging of the metallic island by means of electron tunneling across the insulating barrier as illustrated in Fig.~\ref{fig:fig1}(b). These tunneling transitions may also involve absorption or emission of the resonator photons. Since the rate of the photon-assisted tunneling events and the relative strength between the absorptive and emissive processes is highly dependent on the bias voltage, the voltage-biased SINIS junction provides an effective means for either cooling~\cite{Tan16} or heating~\cite{Masuda16} the resonator. Thus we refer to the SINIS junction and its coupling circuitry as a quantum-circuit refrigerator.

Note that inelastic tunneling processes are familiar from $P(E)$-theory accounting for the energy exchange between a tunneling electron and an electromagnetic environment in thermal equilibrium~\cite{Devoret90,Girvin90,IngoldNazarov05}. In our case, however, the resonator can be driven to a state far from equilibrium.

\begin{figure} 
\includegraphics[width=0.9\linewidth]{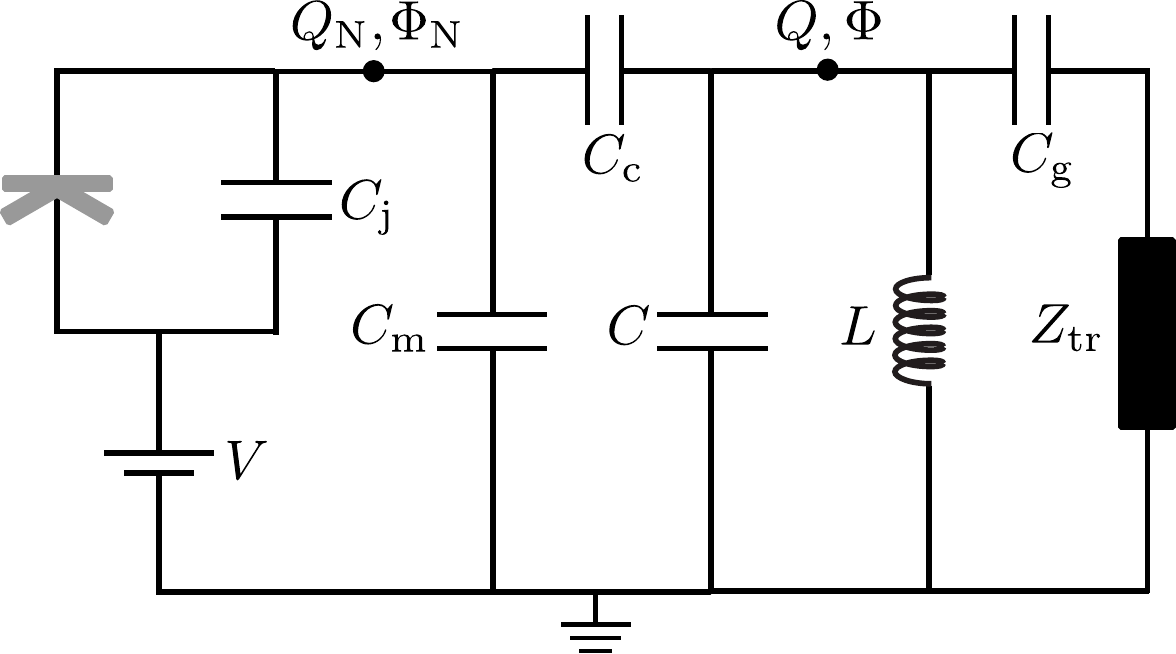}
\caption{\label{fig:fig2}Effective circuit diagram of the studied system. The fundamental mode of the coplanar waveguide resonator is modeled by the lumped-element capacitance~$C$ and inductance~$L$. The resonator couples to a normal-metal island through an input capacitance~$C_{\rm c}$. Tunneling is depicted by the gray symbol on the left representing the weak coupling between a normal-conducting and a superconducting electrode. The diagram shows only one of the NIS junctions with junction capacitance~$C_{\rm j}$ and voltage bias~$V=V_{\rm B}/2$. For tunneling through this junction, the other parallel junction acts as a capacitor, the capacitance of which is included in the capacitance~$C_{\rm m}$ of the metallic island to the ground.  An output capacitor of the capacitance~$C_{\rm g}$ couples the resonator to a transmission line with a characteristic impedance~$Z_{\rm tr}$. The node fluxes at the island and at the resonator are denoted by~$\Phi_{\nmd}$ and~$\Phi$, respectively, and~$Q_{\nmd}$ and~$Q$ are their conjugate charges.}
\end{figure}

The NIS tunnel junctions are assumed to be of high tunneling resistance $R_{\rm T} \sim 10-100$~k$\Omega$. Accordingly, the tunnel coupling can be treated as a weak perturbation and tunneling across each of the junctions can be considered independently. Thus the fundamental mode of the resonator coupled to the quantum-circuit refrigerator can be described by the effective single-junction circuit diagram shown in Fig.~\ref{fig:fig2}. Based on this circuit, we describe below a quantum-mechanical model for the system. 

\section{Model Hamiltonian} \label{sec:ModHamil} 
The core of the system is formed by the coplanar waveguide resonator, the normal-metal island, and the capacitive coupling to the voltage-biased superconducting leads. We first establish the Hamiltonian $\hat{H}_0$ of the core part. Subsequently, we add the weak capacitive coupling to the transmission line and the weak tunnel coupling to the~superconducting~leads.

\subsection{Hamiltonian of the core circuit}
In our analysis below, we utilize the lumped-element circuit of Fig.~\ref{fig:fig2} where we define the various physical capacitances of the system and the inductance $L$ corresponding to the fundamental resonator mode. Following the standard procedure~\cite{Devoret99}, the Lagrangian of the core circuit, $\mathcal{L}_0$, can be expressed in terms of the  fluxes at the two nodes, \textit{i.e.}, the flux $\Phi_{\nmd}$ at the normal-metal island and the flux $\Phi$ at the resonator.
We find
\begin{align}
  \label{eq:lag0}
    \mathcal{L}_0=&\frac{C_{\rm j}}{2}(\dot \Phi_{\nmd}-{V})^2 +\frac{C_{\rm m}}{2} \dot \Phi_{\nmd}^2 +\frac{C_{\rm c}}{2}(\dot \Phi_{\nmd}-\dot \Phi)^2\notag \\ \qquad &+\frac{C}{2} \dot\Phi^2-\frac{\Phi^2}{2L}.
\end{align}
Next we introduce the conjugate charges $Q_{\nmd}=\partial \mathcal{L}_0/\partial \dot \Phi_{\nmd}$ and $Q=\partial \mathcal{L}_0/\partial \dot \Phi$ and the classical Hamiltonian $H_0$ through the Legendre transformation $H_0=Q_{\nmd} \dot \Phi_{\nmd} + Q\dot \Phi-\mathcal{L}_0$. After quantization and neglecting irrelevant constant terms, we obtain the core Hamiltonian operator
\begin{equation}
  \hat H_0 = \frac{(\hat Q_{\nmd}+C_{\rm j}{V})^2}{2 C_{\nmd}}+\frac{[\hat Q+\alpha(\hat Q_{\nmd}+C_{\rm j}{V})]^2}{2C_{\rm r}}+\frac{\hat{\Phi}^2}{2L}, \label{ham.HG}
\end{equation}
where the renormalized capacitances of the island and the resonator are given by
\begin{subequations}\label{cren}
  \begin{align}
    C_{\nmd}&=C_{\rm c}+C_{\Sigma \rm {m}},\\
    C_{\rm r}&=C+\alpha  C_{\Sigma \rm {m}}.
  \end{align}
\end{subequations}
The total capacitance of the normal metal island to ground $C_{\Sigma {\rm m}}=C_{\rm m}+C_{\rm j}$ is independent of the considered junction. The capacitance ratio
\begin{equation}
  \alpha =\frac{C_{\rm c}}{C_{\nmd}}=\frac{C_{\rm c}}{C_{\rm c}+C_{\Sigma \rm{m}}},   \label{eq:cap_fraction}
\end{equation}
characterizes the strength of the capacitive coupling between the normal-metal island and the resonator. Since the applied voltage causes only a constant charge shift $Q_{\rm j}=C_{\rm j}V$ in the Hamiltonian~\eqref{ham.HG}, we can eliminate it with a gauge transformation~\cite{Koch09} where states transform as $\ket{\psi'}=\ee^{\frac{\ii}{\hbar} Q_{\rm j} \hat \Phi_{\nmd}}\ket{\psi}$ and the charge operator as
\begin{equation}
\ee^{\frac{\ii}{\hbar} Q_{\rm j} \hat \Phi_{\nmd}} (\hat Q_{\nmd}+Q_{\rm j}) \ee^{-\frac{\ii}{\hbar} Q_{\rm j} \hat \Phi_{\nmd}}=\hat Q_{\nmd}, \label{eq.gauge}
\end{equation}
owing to the commutation rule $[\hat \Phi_{\nmd}, \hat Q_{\nmd}]=\ii \hbar $. In this gauge, the core Hamiltonian~$\hat H_0$ simplifies to
\begin{equation}
    \hat H_0 = \frac{\hat Q_{\nmd}^2}{2 C_{\nmd}}+\frac{(\hat Q+\alpha  \hat Q_{\nmd})^2}{2C_{\rm r}}+\frac{\hat{\Phi}^2}{2L}, \label{ham.HG1}
\end{equation}
which is independent of the applied voltage. 

\subsection{Coupling to transmission line}
For readout and external control, the core circuit is coupled to a transmission line of characteristic impedance $Z_{\rm tr}\sim 50$ $\Omega$ through an output capacitance $C_{\rm g}$. The impedance of this added element is
\begin{equation}\label{Zom}
Z(\omega)=Z_{\rm tr}+\frac{1}{\ii \omega C_{\rm g}}=Z_{\rm tr}\,\frac{\omega-\ii \omega_{\rm RC}}{\omega},
\end{equation}
where $\omega_{\rm RC}=1/(Z_{\rm tr}C_{\rm g})$ is the inverse of the corresponding $RC$ time. The transmission line acts as an electrodynamic environment of the core circuit. In the limit of weak coupling between the resonator and the transmission line, the influence of the environment is described by the admittance at the resonator frequency $Y(\omega_{\rm r})=Z^{-1}(\omega_{\rm r})$. The real part of $Y(\omega_{\rm r})$ determines the coupling strength, also referred to as the damping coefficient,
\begin{equation}
\gamma_{\rm tr}=\frac{\omega_{\rm r}^2}{C_{\rm r}Z_{\rm tr}(\omega_{\rm r}^2+\omega_{\rm RC}^2)} =\frac{Z_{\rm r}}{Z_{\rm tr}} \frac{\omega_{\rm r}^3}{\omega_{\rm r}^2+\omega_{\rm RC}^2}, \label{eq:gamma}
\end{equation}
where $Z_{\rm r}=\sqrt{L/C_{\rm r}}$ is the characteristic impedance of the resonator. The imaginary part of $Y(\omega_{\rm r})$ leads to a further renormalization of the resonator capacitance by $C'_{\rm r}=C_{\rm r}+\omega_{\rm RC}/[Z_{\rm tr}(\omega^2_{\rm r}+\omega_{\rm RC}^2)]$ and the resonator frequency by
\begin{equation}
\omega_{\rm r}' \approx \omega_{\rm r}-\frac{Z_{\rm r}}{2Z_{\rm tr}}\frac{\omega^2_{\rm r} \omega_{\rm RC}}{\omega_{\rm r}^2+\omega_{\rm RC}^2}.
\end{equation}
Henceforth, we consider the renormalized resonator capacitance and frequency but drop the primes for notational simplicity. 

\subsection{Tunneling Hamiltonian} \label{sec.tun}
To incorporate the weak tunnel coupling between the metallic island and one of the superconducting leads, we need to introduce the quasiparticle degrees of freedom in the electrodes. The electronic Hamiltonian
\begin{equation}\label{Hel}
\hat H_{\rm el}=\hat H_{\nmd}+\hat H_{\rm S}+\hat H_{\rm T},
\end{equation}
is formed by the energy of the conduction electrons in the normal metal island $\hat H_{\nmd}$, the energy of the quasiparticles in the superconductor $\hat H_{\rm S}$, and the energy related to the tunneling interaction $\hat H_{\rm T}$. In the gauge~\eqref{eq.gauge}, we have~\cite{IngoldNazarov05}
\begin{subequations}
\begin{align}
  \hat H_{\nmd}=&\sum_{l\sigma} \varepsilon^{}_{l}\hat d^\dagger_{l\sigma} \hat d^{}_{l\sigma},   \\
  \hat H_{\rm S} =&\sum_{k\sigma}(\epsilon^{}_{k}-eV)\hat c^\dagger_{k\sigma} \hat c^{}_{k\sigma}+\sum_{k}\left(\Delta^{}_k\hat c^\dagger_{k\uparrow} \hat c^\dagger_{-k\downarrow}+\text{h.c.}\right),  \\
   \hat H_{\rm T}=&\sum_{kl\sigma}\left(T^{}_{lk}\hat d^\dagger_{l\sigma}\hat c^{}_{k\sigma}\ee^{-\ii \frac{e}{\hbar} \hat \Phi_{\nmd}}+\text{h.c.} \right),
\label{eq:Ht}
\end{align}
\end{subequations}
where $\varepsilon_{l}$ denotes the energy of the normal-metal quasiparticles with wave vector $l$, spin index $\sigma \in \{\uparrow, \downarrow \}$, and annihilation operator $\hat d^{}_{l \sigma}$. Similarly, for the superconductor quasiparticles, we have the energy $\epsilon_{k}$,  wave vector $k$, and annihilation operator $\hat c^{}_{k\sigma}$. Quasiparticles of the superconductor are coupled through the gap parameter~$\Delta_k$.

 A tunneling event is associated with a change of the quantized charge $\hat Q_{\nmd}$ on the island.  In the tunneling Hamiltonian~$\hat H_{\rm T}$, this is implemented by the operators $\exp(\pm\ii e \hat \Phi_{\nmd}/\hbar)$ where $e$ is the elementary charge~\cite{Devoret90}. Owing to the commutation rule $[\hat \Phi_{\nmd}, \hat Q_{\nmd}]=\ii \hbar $, we have
\begin{equation}
\ee^{\pm\ii \frac{e}{\hbar} \hat \Phi_{\nmd}}\hat Q_{\nmd} \ee^{\mp\ii \frac{e}{\hbar} \hat \Phi_{\nmd}}=\hat Q_{\nmd}\mp e.
\end{equation}
The probability of theses charge transfer processes is proportional to the tunneling matrix elements $T_{lk}$.

Next we transform the voltage bias to the operators by applying a time-dependent unitary transformation $\hat U_V(t) = \prod_{k\sigma} \exp(\ii \frac e \hbar  V t \hat c^\dagger_{k\sigma}\hat c^{}_{k\sigma})$. The Hamiltonians transform according to
\begin{equation}
\hat H'=\hat U_V^\dagger \hat{H} \hat U^{}_V +\ii\hbar (\partial_t \hat U_V^\dagger)  \hat U^{}_V.
\end{equation}
In what follows, we again drop the primed notation for simplicity. The transformation affects the terms $\hat H_{\rm S}$ and $\hat H_{\rm T}$, which become
\begin{subequations}  \label{eq:Htp}
\begin{align}
\hat H_{\rm S}=&\sum_{k\sigma} \epsilon^{}_{k}\hat c^\dagger_{k\sigma} \hat c^{}_{k\sigma}+\sum_{k}\left[\widetilde{\Delta}^{}_k(t)\hat c^\dagger_{k\uparrow} \hat c^\dagger_{-k\downarrow}+\hbox{h.c.}\right], \\
  \hat H_{\rm T}=&\sum_{kl\sigma}\left[T^{}_{lk}\hat d^\dagger_{l\sigma}\hat c^{}_{k\sigma}\ee^{-\ii \frac{e}{\hbar} \left (\hat \Phi_{\nmd}-V t\right)}+\hbox{h.c.}\right].
\end{align}
\end{subequations}
The transformed gap parameter $\widetilde{\Delta}_k(t)=\Delta_k\ee^{-\ii \frac {2 e}{\hbar} V t}$ does not change the superconductor density of states. The charge shift operators of the transformed tunneling Hamiltonian contain a time-dependent phase arising from the applied voltage. The Hamiltonians $\hat{H}_{\nmd}$ for the normal-metal island and $\hat{H}_{0}$ for the core circuit remain unchanged. 

\section{Transition rates from tunnel coupling}~\label{sec:Tunnel}
In this section, we formulate the eigenstates of the core circuit $\hat H_0$ and investigate transitions between them  induced by the tunneling Hamiltonian $\hat H_{\rm T}$. Transition rates corresponding to an NIS junction are derived first and then generalized for the full SINIS junction.

\subsection{Eigenstates of the core circuit}\label{sec:Eigs}
In the previous sections, we introduced the different constituents of the Hamiltonian of the circuit shown in Fig.~\ref{fig:fig2} in the presence of tunnel coupling
\begin{equation}\label{Htot}
\hat H=\hat H_0+\hat H_{\nmd}+\hat H_{\rm S}+\hat H_{\rm T}.
\end{equation}
The coupling to the transmission line is considered separately. In the weak tunneling limit, the term $\hat H_{\rm T}$ can be treated as a small perturbation. In the unperturbed Hamiltonian the parts $\hat H_0$, $\hat H_{\nmd}$, and $\hat H_{\rm S}$ mutually commute, and hence we choose to diagonalize $\hat H_0$.

Since the core Hamiltonian $\hat H_0$ in Eq.~\eqref{ham.HG1} and the island charge operator $\hat Q_{\nmd}$ commute, $[\hat H_0,\hat Q_{\nmd}]=0$, they share eigenstates. First, we denote the charge eigenstates of the normal metal island by $\ket{q}$, where $q$ is an integer such that $\hat Q_{\nmd} \ket{q}=eq\ket{q}$.
Thus the core Hamiltonian may be written as
\begin{equation}
    \hat H_0 = \sum_{q=-\infty}^\infty \sum_{m=0}^\infty \ket{q\,m_q}\bra{q\, m_q} \left[ E_{\nmd} q^2+\hbar \omega_{\rm r} \left(m+\frac 1 2\right)\right],
\end{equation}
where we have introduced the charging energy of the island $E_{\nmd}=e^2/2C_{\nmd}$ and have diagonalized the resonator. The eigenstates of the resonator with frequency $\omega_{\rm r}=1/\sqrt{LC_{\rm r}}$ are harmonic oscillator states,
\begin{equation}
\ket{m_q}=\ee^{-\ii\alpha  q \frac{e}{\hbar} \hat \Phi } \ket{m}, \label{eq.shiftFock}
\end{equation}
where the subscript $q$ indicates the shift of the oscillator coordinate $Q$ by the charge on the normal metal island. For the harmonic oscillator, the charge shift affects only the eigenstates but not the energies. Furthermore, the shifted harmonic oscillator wavefunctions have nontrivial overlap $\braket{m^\prime_{q^\prime}|m^{}_q}$ which can be considered as the origin of the possibility to control the resonator state by photon-assisted tunneling.

\subsection{Tunneling at an NIS junction}
The eigenstates of the core circuit are linked by transitions of electrons from the superconductor to the normal-metal island and vice versa, governed by the tunneling operator
\begin{equation}
\hat \Theta =\sum_{kl\sigma} T^{}_{lk}\hat d^{\dagger}_{l\sigma} \hat c^{}_{k\sigma}.
\end{equation}
The tunneling Hamiltonian~\eqref{eq:Htp} may be expressed as
\begin{equation}
\hat H_{\rm T}=\hat \Theta\, \ee^{-\ii\frac{e}{\hbar}\left(\hat \Phi_{\nmd} - Vt\right)} +\hbox{h.c.}.
\end{equation}
We denote by $\ket E$ an eigenstate of the junction electrodes which is a product of an eigenstate of the normal-metal island $\hat H_{\nmd}$ and that of the superconducting lead $\hat H_{\rm S}$. The relevant transition matrix element is thus of the form
\begin{align}\label{me}
\bra {E^\prime\! ,q^\prime m'_{q^\prime} }&\hat H_{\rm T}\ket {E,q\, m_q}\\
= &  \ee^{+\ii\frac{e}{\hbar}Vt}\braket{E^\prime \vert \hat \Theta \vert E} \braket{q^\prime m^\prime_{q^\prime} \vert \ee^{-i\frac{e}{\hbar}\hat \Phi_{\nmd} } \vert q\, m^{}_q} \notag \\
  & +\ee^{-\ii\frac{e}{\hbar}Vt}\braket{E^\prime \vert  \hat \Theta^\dag \vert E} \braket{q^\prime m^\prime_{q^\prime} \vert \ee^{i\frac{e}{\hbar}\hat \Phi_{\nmd} }  \vert q\, m^{}_q}, \notag
\end{align}
where the matrix elements of the core circuit are given by
\begin{equation}
\braket{q^\prime m^\prime_{q^\prime}\vert \ee^{\pm \ii\frac{e}{\hbar}\hat \Phi_{\nmd} } \vert q\, m^{}_q}=\delta_{q^\prime\! ,q\pm 1}\braket{{m^\prime_{q\pm 1}} \vert m^{}_q}.
\end{equation}
The overlap between the charge-shifted resonator eigenstates can be calculated using Eq.~\eqref{eq.shiftFock}. The result is expressed in terms of the generalized Laguerre polynomials $L^\ell_n(\rho)$.~\cite{Stegun, Hollenhorst79, Catelani11, Souquet14}  We find $|\braket{m^\prime_{q\pm1}|m^{}_q}|^2=|\braket{m^\prime_{\pm1}|m^{}_0}|^2=M_{mm^\prime}^2$ where
\begin{equation}
M_{mm^\prime}^2=\begin{cases}\ee^{-\rho} \rho^\ell \frac{m^\prime!}{m!} \left[L^{\ell}_{m^\prime}(\rho)\right]^2, & m \geq m^\prime, \\ &\\
 \ee^{-\rho} \rho^{-\ell} \frac{m!}{m^\prime!} \left[L^{-\ell}_{m}(\rho)\right]^2, & m <  m^\prime. \end{cases} \label{eq:matrele}
\end{equation}
Above, $\ell=m-m^\prime$ and the interaction parameter
\begin{equation}
\rho=  \pi \alpha ^2  \frac{1}{\omega_{\rm r} C_{\rm r} R_{\rm K}} = \pi \frac{C^2_{\rm c}}{C^2_{\nmd}} \frac{Z_{\rm r}}{R_{\rm K}}, \label{eq.rhoInteraction}
\end{equation}
where $R_{\rm K}=h/e^2$ is the von Klitzing constant defined in terms of the Planck constant $h$ and the elementary charge~$e$. Note that when the interaction is strong, $\rho \gtrsim 1$, some transition overlaps $M^2_{mm^\prime}$ can vanish at the roots of the Laguerre polynomials~\cite{Gramich13}.

To evaluate the electronic matrix elements, such as $\bra {E^\prime} \hat{\Theta} \ket E$, we employ the standard approach for tunnel junctions~\cite{IngoldNazarov05}. Subsequently, we insert the transition matrix element in Eq.~\eqref{me} into Fermi's golden rule evaluated for the Hamiltonians~\eqref{eq:Htp} and trace out the electronic degrees of freedom, and hence obtain the transition rates between eigenstates of $\hat H_0$.

The electronic transitions from the normal-metal island to the superconducting electrode are connected with a transition of the core circuit from a state $\ket {q\, m_q}$ to a state $\ket {q\!+\!1\, m^\prime_{q+1}}$ where the normal-metal island charge has increased by an elementary charge $e$. For these forward-tunneling transitions, we find
\begin{align}
  &\overrightarrow{\Gamma}_{q,m,m^\prime}(V)=\frac{M_{mm^\prime}^2}{e^2R_{\rm T}}
     \iint\D{\epsilon_k}\D{\varepsilon_l}\, n_{\rm S}(\epsilon_k)[1-f_{\rm S}(\epsilon_k)] f_{\nmd}(\varepsilon_l)\notag \\
&\times\delta\left[\epsilon_k+E_{\nmd}(1+2q)+\hbar\omega_{\rm r}(m^\prime-m)-\varepsilon_l-eV\right].\label{eq:ifinteg}
\end{align}
Here, we have assumed that the tunneling matrix elements $|T_{lk}|$ are approximately constant over the relevant integration range around the Fermi energies and have expressed the summation $\sum_{kl \sigma}|T_{lk}|^2$ in terms of the experimentally measurable tunneling conductance $1/R_{\rm T}$. This wide-band limit is appropriate for metallic tunnel junctions. The functions $f_{\rm S}(\varepsilon)$ and $f_{\nmd}(\varepsilon)$ denote the Fermi functions in the superconducting and normal-metal electrodes, respectively. The normalized quasiparticle density of states in the superconductor is given by~\cite{Dynes78}
\begin{equation}
n_{\rm S}(\varepsilon)=\left\vert\textrm{Re}\left\{\frac{\varepsilon+i\gamma_{\rm D}\Delta}{\sqrt{(\varepsilon+i\gamma_{\rm D}\Delta)^2-\Delta^2}}\right\}\right\vert, \label{eq:dos}
\end{equation}
where $\Delta$ is the superconductor gap parameter and the Dynes parameter $\gamma_{\rm D}$ determines the density of subgap states. Typical values for the tunnel junctions are $\gamma_{\rm D}\sim 10^{-4}$ and $\Delta \sim 200$ $\mu$eV ($\Delta/h \sim 50$~GHz). 

It is convenient to introduce the normalized rate of forward quasiparticle tunneling for $R_{\rm T}=R_{\rm K}$ at the energy bias $E$ as
\begin{equation}\label{eq:PEforw}
\normrateright(E) = \frac{1}{h}\int \D{\varepsilon}\, n_{\rm S}(\varepsilon) [1-f_{\rm S}(\varepsilon)]f_{\nmd}(\varepsilon-E).
\end{equation}
In terms of this rate function, the result (\ref{eq:ifinteg}) assumes an intuitive form
\begin{equation}\label{Gammaright1}
\overrightarrow{\Gamma}_{q,m,m^\prime}(V)\!=\! M_{mm^\prime}^2 \frac{R_{\rm K}}{R_{\rm T}}\normrateright{\textbf (}eV+\hbar\omega_{\rm r} \ell
-E_q^+{\textbf )},
\end{equation}
where again $\ell=m-m'$ and
\begin{equation}
E_q^\pm=E_{\nmd}(1\pm 2 q)
\end{equation}
is the change of the charging energy of the normal metal island due to forward/backward tunneling. Likewise, we obtain the rate for the backward tunneling processes related to a transition of the core circuit from a state $\ket{q\, m_q}$ to a state $\ket{q\!-\!1\, m^\prime_{q-1}}$ as
  \begin{equation}\label{Gammaleft1}
\overleftarrow{\Gamma}_{q,m,m^\prime}(V)\!=\! M_{mm^\prime}^2 \frac{R_{\rm K}}{R_{\rm T}} \normrateleft{\textbf (}eV- \hbar\omega_{\rm r}\ell +E_q^- {\textbf )}
  \end{equation}
with the function
\begin{equation}
\overleftarrow{F}(E) =\frac{1}{h}\int \D{\epsilon}\, n_{\rm S}(\varepsilon) f_{\rm S}(\varepsilon)  \left[1-f_{\nmd}\left(\varepsilon-E \right)\right] \label{eq:PEback}
\end{equation}
giving the normalized rate of the backward quasiparticle tunneling events at the energy bias $E$.

\subsection{Symmetric SINIS junction}
The SINIS junction consists of an SIN and an NIS junction. The above results~\eqref{Gammaright1}--\eqref{Gammaleft1} are for the NIS junction. Assuming that the tunneling resistances and junction capacitances are identical, the consideration of the SIN junction is analogous, except that the voltage $V$ is reversed, see Fig.~\ref{fig:fig1}. 
For forward tunneling events across the SIN junction, related to a transition of the core circuit from a state $\ket{q\, m_q}$ to a state  $\ket{q\!+\!1,m^\prime_{q+1}}$, we obtain the forward rate
\begin{equation}\label{Gammaright2}
\overrightarrow{\Gamma}^\prime_{q,m,m^\prime}(V)\!=\overrightarrow{\Gamma}_{q,m,m^\prime}(-V)\!
\end{equation}
provided that the two superconducting electrodes have identical temperatures and densities of states. This symmetry holds also for the backward rate. By summing the contributions from both junctions, the total forward  rate between the states $\ket{q \,m}$ and $\ket{q + 1\,m^\prime}$ is
\begin{equation}
\Gamma^+_{q,m,m^\prime}(V) = \sum_{\tau=\pm 1} \overrightarrow{\Gamma}_{q,m,m^\prime}(\tau V).
\end{equation}
A corresponding result holds for the total backward rate~$\Gamma^-_{q,m,m^\prime}(V)$ between the states $\ket{q \,m}$ and $\ket{q - 1\,m^\prime}$.

To further evaluate the transition rates, we assume that the normal and superconducting electrodes are at the same temperature $T_{\nmd}$. Thus the normalized rate of forward tunneling transitions in Eq.~\eqref{eq:PEforw} may be expressed as
\begin{equation}\label{P+2}
\normrateright(E) =\int \D{\varepsilon}\,n_{\rm S}(\varepsilon) \frac{f(\varepsilon-E)-f(\varepsilon)}{h\left(1-\ee^{-E/k_{\rm B} T_{\nmd }} \right)},
\end{equation}
from which we obtain useful identities $\normrateleft(E) = \normrateright(-E)$ and $\normrateright(-E)=e^{-E/k_{\rm B} T_{\nmd}}\normrateright(E)$ relating the backward and forward rates.
This allows to write the transition rates in both directions fully in terms of the normalized forward rates
\begin{align}
\Gamma^\pm_{q,m,m^\prime}(V)=M^2_{m m^\prime} \frac{R_{\rm K}}{R_\textrm{T}}\Big[&\normrateright \left(eV+\hbar \omega_{\rm r} \ell-E_q^\pm \right)\label{transrates.simp} \\
&+\normrateright \left(-eV+\hbar \omega_{\rm r} \ell-E_q^\pm \right)\Big], \notag 
\end{align}
where $\ell=m-m^\prime$. The only remaining difference between the tunneling directions is in the charging energy differences $E_q^\pm=E_{\nmd}(1\pm 2 q)$.

In practice, the tunneling resistances and capacitances of the two junctions will differ typically by a few percent. The main effect of this asymmetry is a slight shift of the most probable charge state of the island away from the balanced value of $q=0$~\cite{Ingold91b, IngoldNazarov05, Silveri17}. If the charging energy associated with this shift remains small compared with the excitation energy $\hbar\omega_{\rm r}$ of the resonator, which is typically the case, the effects of a modest asymmetry of the rates~\eqref{transrates.simp} are very small. The theory presented here can thus be safely compared with experiments. 
\section{Master equation}~\label{sec:Master}
In the previous sections, we determined the transition rates between the eigenstates of the core circuit induced by a capacitively coupled SINIS junction. The weak coupling to the transmission line induces further transitions, yet, only between neighboring states of the resonator. The rates for these transitions are
\begin{subequations}\label{tlrates}
\begin{align}
{\Gamma}^{\downarrow}_m &=\gamma_{\rm tr}(N_{\rm tr} +1)m,\\
{\Gamma}^{\uparrow}_m &=\gamma_{\rm tr}\, N_{\rm tr} (m+1),
\end{align}
\end{subequations}
where $N_{\rm tr}=1/[\exp(\hbar\omega_{\rm r}/k_{\rm B}T_{\rm tr})-1]$ is the mean thermal occupation factor at the reservoir temperature $T_{\rm tr} \approx 100-150$~mK and the coupling strength to the transmission line $\gamma_{\rm tr}$ is given in Eq.~\eqref{eq:gamma}.

Having determined all transition rates, we write the master equation describing the population dynamics of the core circuit. Let us denote by $p_{q,m}(t)$ the probability that the core circuit occupies the state $\ket{q\,m_q}$ at time $t$. This probability obeys the Pauli master equation
\begin{align}
\dot p_{q,m} = {\Gamma}_{m\!-\!1}^{\uparrow}\,p_{q,m\!-\!1}+{\Gamma}_{m\!+\!1}^{\downarrow}\,p_{q,m\!+\!1}-\left({\Gamma}_{m}^{\uparrow}+{\Gamma}_{m}^{\downarrow}\right)&p_{q,m} \notag \\
+\sum_{m^\prime=0}^\infty \Big[\Gamma^+_{q\!-\!1, m^\prime, m}\,p_{q\!-\!1,m^\prime}+\Gamma^-_{q\!+\!1, m^\prime, m}\, &p_{q\!+\!1,m^\prime}  \notag \\
-\left(\Gamma^+_{q, m, m^\prime}+\Gamma^-_{q, m, m^\prime}\right)p_{q,m^\prime}&\Big]. \label{ME}
\end{align}
We are primarily interested in the steady-state solution of the resonator. Since also the dynamics of the metal island charge $\ket{q}$ is involved, solving the full master equation is a challenging problem. Fortunately, it can be simplified in specials cases as described below.

\subsection{Weak interaction and charge thermalization}
We begin the simplifications of the master equation by considering the magnitude of the resonator matrix elements $M_{mm^\prime}^2$ appearing in the transition rates in Eq.~\eqref{transrates.simp}. The magnitude is governed by the interaction parameter $\rho=\pi \alpha ^2 Z_{\rm r}/R_{\rm K}$ given in Eq.~\eqref{eq.rhoInteraction}. It characterizes how strongly the core circuit responds to a quasiparticle tunneling event. The ratio of the characteristic impedance of the resonator $Z_{\rm r}$ and the von Klitzing constant $R_{\rm K}$ is a measure for the stiffness of the harmonic oscillator potential, determining how strongly a charge shift affects the overlap of the resonator eigenstates.  The magnitude of the charge shift, in turn, depends on the capacitance ratio $\alpha =C_{\rm c}/C_{\nmd}$. The capacitive coupling can be made strong~\cite{Masuda16} by choosing in the fabrication the input capacitance $C_{\rm c}$ to dominate over the normal metal capacitance $C_{\Sigma \rm m}$ in $C_{\nmd}=C_{\rm c}+C_{\rm \Sigma m}$. In fact, a galvanic contact~\cite{Tan16} realizes $\alpha =1$. Despite of strong capacitive interaction, typical geometries and material parameters of coplanar waveguide resonators render the characteristic impedance low compared to $R_{\rm K}$. Hence, with resonators, the interaction parameter has typically very low values of the order of $\rho\sim 0.001$. 

\begin{figure}
  \includegraphics[width=1.0\linewidth]{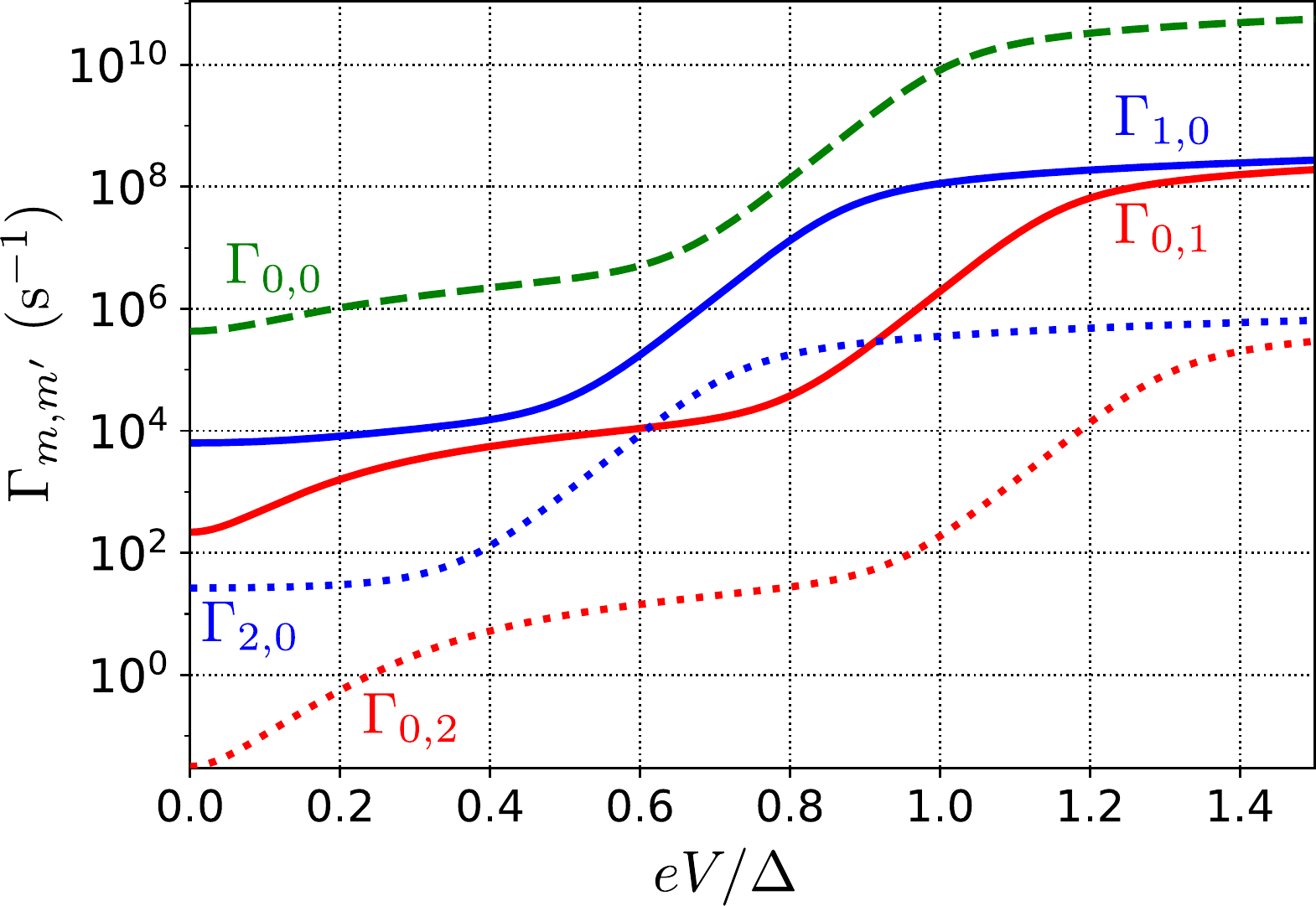}
  \caption{\label{fig:gammaN} Resonator transition rates ${\Gamma}_{m,m^\prime}$ of Eq.~\eqref{Gammares} as functions of the single-junction bias voltage. Both the one-photon processes ${\Gamma}_{1,0}$ (blue solid line) and ${\Gamma}_{0,1}$ (red solid line) as well as the two-photon processes ${\Gamma}_{2,0}$ (blue dotted line) and ${\Gamma}_{0,2}$ (red dotted line) are shown. The elastic transition rate ${\Gamma}_{0,0}$ (green dashed line) exceeds the inelastic rates.  The used parameters correspond to a typical experiment~\cite{Tan16, Masuda16}: $\Delta=200$~$\mu$eV, $\gamma_{\rm D}=10^{-4}$, $R_{\rm T}=50$~k$\Omega$,  $T_{\nmd}=100$~mK, $C_{\rm c}=1.0$~pF, $C_{\Sigma m}=10$~fF, $\omega_{\rm r}/2\pi = 7.0$~GHz, and $Z_{\rm r}=35\ \Omega$.}
\end{figure}

Let us express the matrix elements $M_{mm^\prime}^2$ in Eq.~\eqref{eq:matrele} in the lowest order in $\rho$. Using
$\ell=m-m^\prime$ and
 $L_{m^\prime}^\ell(0)=\binom{m}{m^\prime}$, we obtain
\begin{equation}
M_{mm^\prime}^2=\begin{cases} \frac{1}{\ell!}\binom{m}{\ell}\rho^{\ell} +\mathcal O \left(\rho^{\ell+1}\right), & m \ge m^\prime, \\
\frac{1}{|\ell|!}\binom{m^\prime}{|\ell|} \rho^{|\ell|}+\mathcal O \left(\rho^{|\ell|+1}\right), & m<m^\prime.
\end{cases} \label{eq:M:lin}
\end{equation}
The small interaction parameter will imply that tunneling processes involving simultaneously several photons are suppressed as $M^2_{m,m\pm\ell} \propto \rho^\ell$ with respect to single-photon processes where $M^2_{m,m \pm 1} \propto \rho$. Despite of this suppression of the matrix elements, we note that  photon-assisted tunneling drives multi-photon transitions, formally, without typical selection rules. In certain conditions, due to the enhancement by electron tunneling rates, the rate of absorptive two-photon transitions can exceed the emissive single-photon rate as shown in Fig.~\ref{fig:gammaN}. In this kind of situations, the resonator state needs to be solved from the master equation~\eqref{ME}, which takes into account all single and multi-photon processes. In this paper, however, we do not concentrate on the details of the high-photon-number processes.

We observe from the master equation~\eqref{ME} that the charge state $\ket{q}$ is driven both by elastic and inelastic transitions. As in the case of multi-photon transitions, the single-photon inelastic transitions are suppressed with respect to elastic transitions since $M^2_{m,m} \propto 1-\rho$. Thus for typical experimental parameters, \textit{i.e.}, when the ratio $\hbar\omega_{\rm r}/\Delta$ is sufficiently small, the temperature low $k_{\rm B}T_{\nmd} \ll \Delta$, and the Dynes parameter $\gamma_{\rm D}$ rather large, the inelastic tunneling rates remain lower than the elastic rates in the whole relevant bias region (see Fig.~\ref{fig:gammaN}). As shown in Appendix~\ref{app:chargethermo}, in the case of dominating elastic tunneling, the charge states of the normal-metal island rapidly approach a stationary and symmetric thermal distribution $p_q=\exp(-E_{\nmd} q^2/k_{\rm B}T_{\rm Q})/Z$, where $T_{\rm Q}$ is the effective temperature of the charge distribution and $Z=\sum_q \exp(-E_{\nmd} q^2/k_{\rm B}T_{\rm Q})$. 
\begin{figure}
  \centering
  \includegraphics[width=0.95\linewidth]{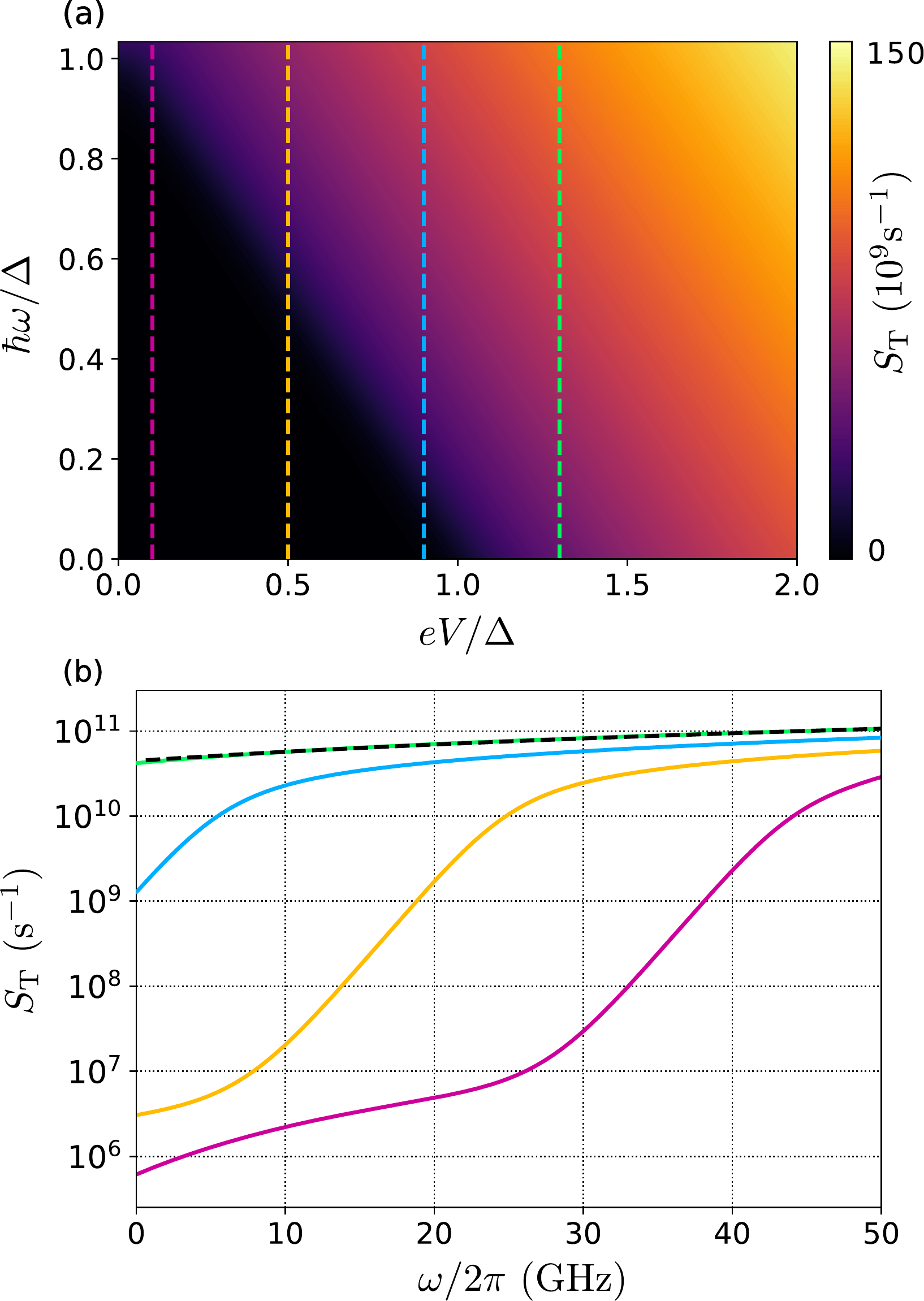}
  \caption{\label{fig:S}(a) Spectral density of tunneling $S_{\rm T}$ as a function of the angular frequency $\omega$ and single-junction bias voltage $V$. The parameters equal to those~of~Fig.~\ref{fig:gammaN}. (b)~Traces from panel~(a) corresponding to $eV/\Delta=0.1, 0.5, 0.9, 1.3$ [vertical dashed lines in panel~(a)]. The black dashed line is a fit to an ohmic behavior $S_{\rm T}\propto \omega$ at $eV/\Delta=1.3$. Near the gap voltage $ |eV-\Delta| \lesssim  \hbar \omega $, the spectral density exhibits clear non-ohmic, exponential behavior. In the region $ |eV-\Delta| \gtrsim \hbar \omega $, the spectral density $S_{\rm T}(\omega)$ flattens to a linear, ohmic trend. }
\end{figure}

\subsection{Master equation for the resonator}
The resonator states $\ket{m}$ are driven by inelastic processes or by transitions induced by the coupling to the transmission line. This latter relaxation is practically independent of the metal island charge. In the case where the elastic tunneling events dominate over the inelastic ones, the charging dynamics of the metal island is thermalized rapidly with respect to the time scales of the inelastic transitions. In an experimentally relevant regime, it suffices to consider the charge and resonator dynamics independently and it is justified to average the resonator transition rates in Eq.~\eqref{transrates.simp} over the stationary and symmetric thermal distribution of the charge states~$p_q$. Consequently, we define
\begin{equation}
{\Gamma}_{m,m^\prime}(V)=\sum_{q} p_q\left[\Gamma^+_{q,m,m^\prime}(V)+\Gamma^-_{q,m,m^\prime}(V)\right] \label{Gammaud}.
\end{equation}
With typical experimental parameters, the charging energy of the normal-metal island is much smaller than other relevant energy scales of the setup: $E_{\nmd}=e^2/2C_{\nmd}\ll \Delta, \hbar\omega_{\rm r}, k_{\rm B}T_{\nmd}$ ($E_{\nmd}/h \sim 10$ MHz). Thus, we expand the transition rates in Eq.~\eqref{Gammaud} to first order in $E_q$ and average over the thermal charge state distribution. Owing to the symmetry of the charge distribution, the first order effect of the charging energy differences $E_q^\pm=E_{\nmd}(1\pm 2q)$ vanishes except for a small overall bias shift by the charging energy $E_{\nmd}$. Hence, the resonator transition rate from the state $\ket{m}$ to the state $\ket{m^\prime}$ assumes the form
\begin{align}
\Gamma_{m,m^\prime}(V) & \approx M^2_{m m^\prime} \frac{2R_{\rm K}}{R_{\rm T}}\sum_{\tau=\pm 1 } \normrateright \left(\tau eV+\hbar \omega_{\rm r} \ell-E_{\nmd} \right)\notag \\
&= M^2_{m m^\prime} S_{\rm T}(\omega_{m m^\prime}). \label{Gammares}
\end{align}
Here, $S_{\rm T}(\omega)$ denotes the spectral density of tunneling and $\omega_{m m^\prime}=\omega_{\rm r}(m-m^\prime)$ the transition frequency. The transition rates and the spectral density are shown in Figs.~\ref{fig:gammaN}~and~\ref{fig:S}, respectively, with typical experimental parameters. Finally, we express the master equation for the population of the resonator states as
\begin{align}\label{PME:res}
\dot p_{m} =  {\Gamma}_{m\!-\!1}^{\uparrow}\,p_{m\!-\!1} & +{\Gamma}_{m\!+\!1}^{\downarrow}\,p_{m\!+\!1}  -\left({\Gamma}_{m}^{\uparrow}+{\Gamma}_{m}^{\downarrow}\right) p_{m} \notag \\
&+\sum_{m^\prime}{\Gamma}_{m^\prime,m}\,p_{m^\prime}  -\sum_{m^\prime} {\Gamma}_{m,m^\prime}\, p_{m}\text{.}
\end{align}
Here, the terms containing $\Gamma_m^{\updownarrows}$ describe the transitions caused by the transmission line and the remaining ones those due to photon-assisted tunneling.

\subsection{Single-photon thermal reservoir}
\begin{figure} 
  \centering
  \includegraphics[width=0.95\linewidth]{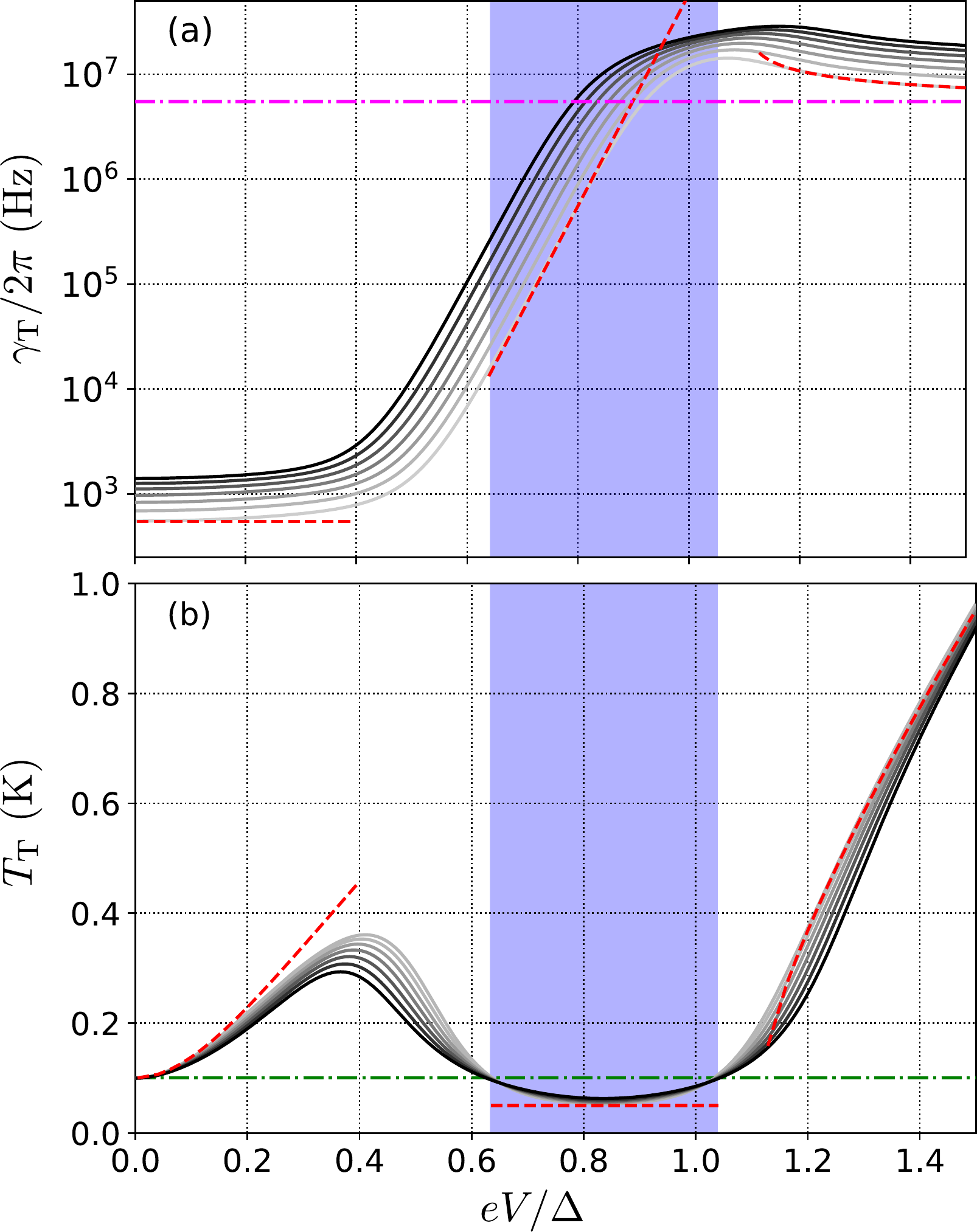}
  \caption{\label{fig:gammaT} (a)~The coupling strength $\gamma_{\rm T}$ and (b) the temperature~$T_{\rm T}$ of the thermal reservoir by photon-assisted tunneling as a function of the single-junction bias voltage. The parameters equal to those of Fig.~\ref{fig:gammaN} except that the resonator frequencies are $\omega_{\rm r}/2\pi=4.0, 5.0, \ldots, 10.0$~GHz corresponding to line colors from light gray to black, respectively. The overall scaling of $\gamma_{\rm T}$ comes from the scaling of $\bar{\gamma}_{\rm T}$ introduced in Eq.~\eqref{eq:gammaTbar} assuming that the characteristic impedance $Z_{\rm r}$ is frequency independent. The coupling strength $\bar{\gamma}_{\rm T}$ for $\omega_{\rm r}/2\pi=4.0$~GHz is depicted by a magenta dash-dotted line. In the highlighted region the thermal reservoir is cooler than the island electrons with temperature $T_{\nmd}$, depicted by a green dash-dotted line. The red dashed lines show the analytic results for $\omega_{\rm r}/2\pi=4.0$~GHz in the deep subgap, thermal activation and above the gap regions, respectively, corresponding to the analytical results~\eqref{eq.deep.gN}, \eqref{eq.gap}, and \eqref{PE:above}.}
\end{figure}

 For the single-photon processes, we observe that $M_{m,m-1}^2=\rho m$, $M_{m,m+1}^2=\rho(m+1)$ which are exactly of the form of the matrix elements of the destruction $\hat a$ and creation $\hat a^\dagger$ operators of the resonator, respectively. Since the single-photon matrix elements $M_{m m \pm 1}^2$ are linear in the photon number $m$, the transition rates in Eq.~\eqref{Gammares} can be written in an intuitive form~\footnote{The coupling strength $\gamma_{\rm T}$ and the mean thermal occupation $N_{\rm T}$ can be formally solved from Eq.~\eqref{Gammares} as $\gamma_{\rm T}= {\bf \Gamma}_{m, m-1}/m-{\bf \Gamma}_{m, m+1}/(m+1)$ and $N_{\rm T}=\Gamma_{m,m+1}/[(m+1)\gamma_{\rm T}]$.}
\begin{subequations} \label{eq:singlerates}
\begin{align}
{\Gamma}_{m, m-1}&=\gamma_{\rm T} (N_{\rm T}+1) m,\\
{\Gamma}_{m, m+1}&=\gamma_{\rm T} N_{\rm T} (m+1).
\end{align}
\end{subequations}
These rates correspond exactly to those generated by coupling to a thermal reservoir with the coupling strength~$\gamma_{\rm T}$ and the mean thermal occupation~$N_{\rm T}$. The effective temperature $T_{\rm T}$ of the reservoir is defined by $N_{\rm T}=1/\left[\exp(\hbar \omega_{\rm r}/k_{\rm B}T_{\rm T})-1\right]$. The subscript T indicates that the reservoir stems from the inelastic tunneling processes either absorbing or emitting resonator photons. The coupling strength and temperature can be written as 
\begin{subequations} \label{eq:gammaN}
\begin{align}
  \gamma_{\rm T} &= \bar\gamma_{\rm T} \frac{\pi}{\omega_{\rm r}}\sum_{\ell,\tau=\pm 1} \ell\normrateright \left(\tau eV+\ell\hbar \omega_{\rm r}-E_{\nmd} \right),  \label{eq:gammaa}\\
T_{\rm T} &= \frac{\hbar\omega_{\rm r}}{k_{\rm B}}\left[ \ln \left( \frac{ \sum_{\tau=\pm 1}  \normrateright \left(\tau eV+\hbar \omega_{\rm r}-E_{\nmd} \right)}{\sum_{\tau =\pm 1}  \normrateright \left(\tau eV-\hbar \omega_{\rm r}-E_{\nmd} \right)}\right) \right]^{-1},  \label{eq:T}
\end{align}
\end{subequations}
 where, for the sake of notational brevity, we have defined
\begin{equation}
\bar{\gamma}_{\rm T}=2\frac{C_{\rm c}^2}{C_{\nmd}^2}\frac{1}{R_{\rm T}C}= 2\frac{C_{\rm c}^2}{C_{\nmd}^2}\frac{Z_{\rm r}}{R_{\rm T}}  \omega_{\rm r}, \label{eq:gammaTbar}
\end{equation}
which is the coupling strength corresponding to a normal-metal--insulator--normal-metal junction. It is determined by the inverse of the $RC$ time for the two parallel junctions each with tunneling resistance $R_{\rm T}$ multiplied by the capacitive coupling fraction $C^2_{\rm c}/C^2_{\nmd}$ of the normal-metal island. Note that the coupling strength $\bar{\gamma}_{\rm T}$ has a weaker dependence on the resonator frequency than the coupling strength to the transmission line of Eq.~\eqref{eq:gamma} for $\omega_{\rm RC}\gg \omega_{\rm r}$. In terms of the effective temperature $T_{\rm T}$ the single-photon rates obey a detailed balance condition
\begin{equation}
\frac{{\Gamma}_{10}}{{\Gamma}_{01}}=\frac{S_{\rm T}(\omega_{\rm r})}{S_{\rm T}(-\omega_{\rm r})}=\ee^{\frac{\hbar \omega_{\rm r}}{k_{\rm B} T_{\rm T}}}.
\end{equation}
In Fig.~\ref{fig:gammaT}, we show the bias voltage dependence of the coupling strength and the effective temperature for typical experimental parameters.

 In this regime where the multiphoton processes are negligible, the quantum dynamics of the circuit in Fig.~\ref{fig:fig2} corresponds to a harmonic oscillator which is linearly coupled to two thermal baths: one arising from the transmission line and the other from the photon-assisted tunneling. The coupling strengths and mean thermal occupations of the baths are known. This model allows us to directly generalize~\cite{Alicki77} the master equation~\eqref{PME:res} to that of the density matrix $\hat \rho$ of a linear resonator
\begin{align}
  \dot{\hat\rho}=&-\frac{\ii}{\hbar}[\hat H_{\rm r}, \hat \rho]\label{ME:res} \\&+\mathcal{D}\left(\sqrt{\gamma_{\rm tr}(N_{\rm tr}+1)}\hat a \right)\hat \rho+\mathcal{D} \left(\sqrt{\gamma_{\rm tr}N_{\rm tr}}\hat a^\dag\right)\hat \rho \notag \\
&+\mathcal{D}\left(\sqrt{\gamma_{\rm T}(N_{\rm T}+1)}\hat a\right)\hat \rho+\mathcal{D}\left(\sqrt{\gamma_{\rm T}N_{\rm T}}\hat a^\dag \right)\hat \rho, \notag
\end{align}
where $\mathcal{D}(\hat c)\hat{\rho}=\hat c \hat \rho \hat c^\dag - \frac 1 2 (\hat c^\dag\hat c \hat \rho+ \hat \rho \hat c^\dag\hat c)$ denotes a Lindbladian dissipator and $\hat H_{\rm r}=\hbar \omega_{\rm r}(\hat a^\dagger \hat a+\frac{1}{2})$ is the Hamiltonian of the resonator. Since tunneling has no effect on the resonator energy, there is no added dephasing of the resonator in Eq.~\eqref{ME:res}. The resulting steady-state occupation of the resonator is a weighted sum of the mean thermal occupation numbers
  \begin{equation}
    N_{\rm r}=\frac{\gamma_{\rm tr} N_{\rm tr}+\gamma_{\rm T} N_{\rm T}}{\gamma_{\rm tr}+\gamma_{\rm T}},\label{eq:Nr}
  \end{equation}
and the total damping coefficient of the resonator is a direct sum the two damping coefficients
  \begin{equation}
    \gamma_{\rm r}=\gamma_{\rm tr}+\gamma_{\rm T}. \label{eq:gr}
  \end{equation}
Hence, the steady-state occupation of the resonator depends both on the relative temperatures and coupling strengths of the two thermal baths acting on the resonator.

The coupling strength $\gamma_{\rm T}$ and the temperature $T_{\rm T}$ inherit the strong bias voltage dependence of the resonator transition rates originating from the tunneling processes, see Fig.~\ref{fig:gammaT}. This is one of the most important results of our work. In the remainder, we will elaborate on these findings, give analytical results in parameter regions of experimental interest, and compare those with experimental data.

Tuning the bias voltage near the edge of the superconductor gap, absorptive transitions become energetically favorable since emissive transitions suffer from the suppressed density of states of the superconductor, see Fig.~\ref{fig:fig1}(b). This implies a cold effective thermal reservoir, ideally $T_{\rm T} = T_{\nmd}/2$. Near the gap, also the coupling strength increases exponentially together with the rate of absorptive forward transitions, see Eq.~\eqref{eq:gammaa}. In other words, near the gap one can achieve a strong coupling to a cold bath, which is the operating principle of the quantum-circuit refrigerator~\cite{Tan16}. When tuning beyond the superconductor gap $eV\gg \Delta$, the relative difference of absorptive and emissive transitions asymptotically vanishes implying a thermal reservoir approaching an infinite temperature. This operating regime can be utilized as incoherent on-chip microwave source as recently demonstrated~\cite{Masuda16}. The coupling strength saturates as the absolute difference between the rates approaches a constant. 

\section{Regions of experimental interest} \label{sec:ExpLimits}
Let us consider in greater detail three parameter regions of particular interest to the recent experimental studies~\cite{Tan16, Masuda16}. First, the bias region $|eV\pm\hbar\omega_{\rm r}|\ll \Delta$, referred to as the deep subgap, where the tunneling is dominated by the remaining superconductor subgap states characterized by the Dynes parameter $\gamma_{\rm D}$ in Eq.~\eqref{eq:dos}. The second region is referred to as the thermal-activation region, where the thermal excitations over the superconductor gap dominate over the subgap transitions. Typically, thermal activation becomes the most important type of transitions for $\Delta/2 \lesssim eV- \hbar \omega_{\rm r}$ and $eV+\hbar \omega_{\rm r} \lesssim \Delta$, where the lower bound $\Delta/2$ depends on the electron temperature of the metals and the Dynes parameter. If the junction is biased above the superconductor gap $\Delta \lesssim eV- \hbar \omega_{\rm r}$, neither subgap states nor thermal excitations are the dominant source of excitations since electrons can tunnel through the insulating barrier just given the energy from the bias voltage. In the extreme biasing regime $eV\gg \Delta$ the shape of the superconductor density of states loses its significance. Since the matrix elements in Eq.~\eqref{eq:matrele} are independent of the bias voltage, we may analytically calculate the tunneling rates in Eq.~\eqref{eq:singlerates}. In the analytic considerations we ignore the charging energy $E_{\nmd}$ since with typical experimental parameters~\cite{Tan16, Masuda16} $E_{\nmd}/\Delta \sim 10^{-3}$.

\subsection{Deep subgap}\label{sec:deepgap}
In the deep-subgap region at small bias, the forward $\normrateright (E)$ and backward tunneling $\normrateright(-E)$ rates are of comparable magnitude. We use the expression~\eqref{P+2} and linearize $\normrateright (E)$ for small $E$ resulting in
\begin{align}
  \normrateright(E) &\approx \frac{\frac{E}{k_{\rm B}T_{\nmd}}}{4h \left(1-\ee^{-\frac{E}{k_{\rm B}T_{\nmd}}}\right)} \int \D{\epsilon}\ \textrm{sech}^2 \left(\frac{\varepsilon}{2k_{\rm B}T_{\nmd}} \right) n_{\rm S}(\epsilon) \notag\\ & \approx \frac{\gamma_{\rm D}E} {h \left(1-\ee^{-\frac{E}{k_{\rm B}T_{\nmd}}}\right)}.
\end{align}
Here the function $\textrm{sech}^{2}[\epsilon/(2k_{\rm B}T_{\nmd})]$ is peaked near the origin with width $\sim k_{\rm B}T_{\nmd}\ll \Delta$. Thus, in the deep-subgap region, the superconductor density of states of Eq.~\eqref{eq:dos} can be approximated by the constant $n_{\rm s}(\epsilon)\approx \gamma_{\rm D}$. Similar arguments apply for the backward tunneling, resulting in the deep-subgap coupling strength $\gamma^{\rm dg}_{\rm T}$ and temperature $T^{\rm dg}_{\rm T}$ to be:
\begin{subequations}\label{eq.deep.gN}
\begin{align}
  &\gamma^{\rm dg}_{\rm T}(V)=2\frac{C^2_{\rm c}}{C_{\nmd}^2} \frac{Z_{\rm r}}{R_{\rm T}} \omega_{\rm r} \gamma_{\rm D}=  \bar{\gamma}_{\rm T} \gamma_{\rm D}, \label{eq:deep_gamma}\\
&T^{\rm dg}_{\rm T}(V)=\frac{\hbar\omega_{\rm r}}{k_{\rm B}}\label{eq:deep_T}\\\notag  \times& \left\{ \log \left[ \frac{\frac{eV}{\hbar \omega_{\rm r}}\sinh\left( \frac{\hbar \omega_{\rm r}}{k_{\rm B} T_{\nmd}} \right)+\cosh \left( \frac{\hbar \omega_{\rm r}}{k_{\rm B} T_{\nmd}} \right)-\ee^{\frac{\hbar \omega_{\rm r}}{k_{\rm B} T_{\nmd}} }} {\frac{eV}{\hbar \omega_{\rm r}}\sinh\left( \frac{\hbar \omega_{\rm r}}{k_{\rm B} T_{\nmd}} \right)-\cosh \left( \frac{\hbar \omega_{\rm r}}{k_{\rm B} T_{\nmd}} \right)+\ee^{-\frac{\hbar \omega_{\rm r}}{k_{\rm B} T_{\nmd}} }}\right]\right\}^{-1}
\end{align}
\end{subequations}
See Eq.~\eqref{eq:gammaTbar} for the definition of $\bar{\gamma}_{\rm T}$. Note that at zero bias, the temperature of the reservoir formed by the SINIS tunnel junction equals to the electron temperature of the metals $T_{\rm T}^{\rm dg}(0)=T_{\nmd}$.

\subsection{Thermal activation}
Beyond the deep-subgap region, the thermally excited electron tunneling starts to dominate. Formally, this means that the tails of the Fermi distributions reach beyond the gap of the superconductor density of states. In this region, for an analytical treatment we ignore the effects of the backward tunneling and the subgap states. Approximating the Fermi function tails by exponentials results in
\begin{align}
\normrateright(E)&\approx \frac{1}{h}\int_{\Delta}^\infty\D{\varepsilon}\: n_{\rm S}(\varepsilon)\exp\left(-\frac{\varepsilon-E}{k_{\rm B}T_{\nmd}}\right)\notag \\
& =\frac{\Delta}{h}  \exp\left(\frac{E}{k_{\rm B}T_{\nmd}}\right) K_1\left(\frac{\Delta}{k_{\rm B}T_{\nmd}}\right)\notag \\ &\approx \frac{1}{h}\sqrt{\frac{\pi k_{\rm B} T_{\nmd} \Delta}{2}} \exp\left(\frac{E-\Delta}{k_{\rm B}T_{\nmd}}\right). \label{eq:Pforw:th}
\end{align}
Here, $K_1[\Delta/(k_{\rm B}T_{\nmd})]$ denotes the modified Bessel function of the second kind~\cite{Stegun}, which is exponentially decaying for $\Delta \gg  k_{\rm B}T_{\nmd}$. Thus the thermally activated coupling strength and the temperature of the reservoir are given by
\begin{subequations}\label{eq.gap}
  \begin{align}
   & \gamma^{\rm th}_{\rm T}(V)=\bar{\gamma}_{\rm T} \frac{\sinh\left(\frac{\hbar \omega_{\rm r}}{k_{\rm B}T_{\nmd}}\right)}{\frac{\hbar \omega_{\rm r}}{k_{\rm B}T_{\nmd}}} \sqrt{\frac{2\Delta}{\pi k_{\rm B}T_{\nmd}}} \exp\left({\frac{e V-\Delta}{k_{\rm B}T_{\nmd}}}\right), \label{eq.gap_gamma}\\
   &T^{\rm th}_{\rm T}(V)=\frac {T_{\nmd}}{2}. \label{eq.gap_T}
  \end{align}
\end{subequations}
Importantly, the effective temperature of the reservoir formed by the SINIS junction is constant and equals to half of the electron temperature. This can be understood by the fact that the tunneling rates for the photon absorption and emission events are weighted by the normal-metal Fermi functions evaluated at the energy difference $2\hbar \omega_{\rm r}$. Yet each of these events causes an energy change of $\hbar \omega_{\rm r}$ in the resonator.

However, the coupling strength is exponentially dependent on the distance of the bias from the superconductor gap. This originates from the fact that the thermally activated transition rates increase exponentially as the effective bias $eV\pm \hbar \omega_{\rm r}$ approaches the gap~$\Delta$. This phenomenon also yields the above exponential dependence of the rates on the ratio $\hbar\omega_{\rm r}/k_{\rm B}T_{\nmd}$. The weak overall scaling of the coupling strength with the square root of the ratio of the superconductor gap to thermal energy $\sqrt{\Delta/k_{\rm B}T_{\nmd}}$ is attributed to the peak in the superconductor density of states near the gap edge.

\subsection{Above the gap}
Above the gap $E=eV\pm\hbar\omega_{\rm r}\gtrsim \Delta$, the dominant source of resonator transitions are the direct photon-assisted tunneling events in the forward direction from an occupied state in the normal metal to an empty state in the superconductor. In this range, backward, subgap and thermal tunneling events are negligible. Using the Sommerfeld expansion in Eq.~\eqref{P+2} results in
\begin{align}
  \normrateright\left(E\right) \approx& \frac{\int_{\Delta}^{E} \D{\varepsilon}\, n_{\rm S}(\varepsilon) + \frac{(\pi k_{\rm B}T_{\nmd})^2}{6} \frac{\D{n}_{\rm S}(\varepsilon)}{\D{\varepsilon}}\big|_{\varepsilon=E}}{h \left(1-\ee^{-E/k_{\rm B}T_{\nmd}} \right)} \notag \\
= &\frac{\sqrt{E^2-\Delta^2}+ \frac{(\pi k_{\rm B}T_{\nmd})^2}{6}\frac{\Delta^2}{(E^2-\Delta^2)^{3/2}}}{h \left(1-\ee^{-E/k_{\rm B}T_{\nmd}} \right)} \label{PE:above}
\end{align}
where $\Delta \gg k_{\rm B}T_{\nmd}$ and the Dynes parameter is ignored. At the asymptotic limit of large bias $eV\gg \Delta, \hbar \omega_{\rm r}$, the resulting coupling strength and the effective temperature of the reservoir become
\begin{subequations}\label{eq:above}
 \begin{align}
\gamma^{\rm ag}_{\rm T}(V)\approx &\bar{\gamma}_{\rm T} \left[1 + \frac{\pi^2}{2} \frac{\Delta^2  k^2_\textrm{B} T^2_{\nmd}}{ (eV)^4} \right],\\
T^{\rm ag}_{\rm T}(V) \approx & \frac{eV}{2k_{\rm B}}\left[1-\frac{2\pi^2}{3} \frac{\Delta^2 k^2_\textrm{B} T^2_{\nmd}}{(eV)^4} \right].
\end{align}
\end{subequations}
Asymptotically, up to the corrections from the thermal broadening, the coupling strength equals that of a fully normal-metal junction, $\bar{\gamma}_{\rm T}$, and the reservoir temperature increases linearly as $eV/(2k_{\rm B})$.

\section{Quantum-circuit refrigerator} \label{sec:QCRcomp}
Here, we discuss how the SINIS junction can be used as a voltage-tunable refrigerator of the resonator mode. In an ideal situation, the inverse of the coupling strength $1/\gamma_{\rm T}$ is the time scale in which the refrigerated quantum circuit exponentially reaches the temperature of the refrigerator $T_{\rm T}$. Thus, to optimize the operation of the quantum-circuit refrigerator, one aims at maximizing the coupling strength $\gamma_{\rm T}$ and reaching the minimal temperature $T_{\rm T}=T_{\nmd}/2$ when the refrigerator is active. To this end, it is instructive to study the coupling strength in the thermal activation regime, \textit{i.e.}, at junction bias near and somewhat below the superconductor gap, where the thermal reservoir formed by the photon-assisted tunneling reaches its lowest effective temperature and is yet strongly coupled to the resonator mode. The  value of the coupling strength at the minimal temperature is of the order of $\bar{\gamma}_{\rm T}=  2 \alpha^2 Z_{\rm r} \omega_{\rm r} / R_{\rm T}$ obtained from Eq.~\eqref{eq:gammaTbar} when biased near the superconductor gap, see Fig.~\ref{fig:gammaT}. This value can be increased by reducing $R_{\rm T}$,  increasing the capacitance fraction $\alpha = C_{\rm c}/C_{\nmd}$, the characteristic impedance $Z_{\rm r}$, or the frequency $\omega_{\rm r}$ of the resonator mode. 

Fortunately, the other optimization goal, $T_{\rm T}\approx T_{\nmd}/2$, is typically achieved in a range of bias voltages where the coupling strength is relatively close to its maximum value, see Fig.~\ref{fig:gammaT}. Note, that a reduction of the electron temperature $T_{\nmd}$ has two positive effects for the quantum-circuit refrigerator. Naturally, it lowers the minimal temperature, but it also increases $\gamma_{\rm T}$ in the thermal activation regime. The increase occurs through the dependence on $\sinh(\hbar \omega_{\rm }/k_{\rm B} T_{\nmd}) \sqrt{k_{\rm B}T_{\nmd}}  / \hbar \omega_{\rm r}$ in Eq.~\eqref{eq.gap}. When the electron temperature is decreased, the minimum reservoir temperature is achieved with smaller bias. When this shift is taken into account, the coupling strength $\gamma_{\rm T}$ at the optimal biasing point exhibits an exponential increase at lowered electron temperatures. Lower electron temperature leads also to a wider thermal activation regime being beneficial for reaching the minimal $T_{\rm T}=T_{\nmd}/2$. Furthermore, reduction of the Dynes parameter $\gamma_{\rm D}$ suppresses subgap transitions which also widens the thermal activation regime. 

In addition to optimization of the cooling rate, the device needs to be well decoupled when not in use. We observe from Fig.~\ref{fig:gammaT}(a) that at $V=0$ the coupling strength reduces to the value $\gamma_{\rm T}^{\rm min}=\bar{\gamma}_{\rm T}\gamma_{\rm D}$. Typically $\gamma_{\rm T}^{\rm min}/2\pi \sim 1$~kHz which is much lower than the typical coupling strengths used to couple coplanar waveguide resonators to transmission lines. Thus the refrigerator becomes essentially decoupled from the resonator. The ratio between the maximum and minimum coupling strength scales as $\gamma^{\rm max}_{\rm T}/\gamma_{\rm T}^{\rm min} \sim \gamma_{\rm D}^{-1}$ and thus it is maximized by reducing the Dynes parameter~$\gamma_{\rm D}$.

\section{Cryogenic microwave source}\label{sec:heatingcomp}
If the junctions are biased above the superconductor gap, the suppression of the emissive transitions due to the gap is lifted. Thus the thermal reservoir describing the photon-assisted tunneling events is hot. In this regime, the device can work as a microwave source emitting incoherent radiation through the resonator at frequency $\omega_{\rm r}$ and a spectral peak width of $\gamma_{\rm tr}+\gamma_{\rm T}$.~\cite{Masuda16}

Let us quantitatively investigate the net power emitted from the capacitively coupled SINIS junction to the transmission line through the resonator. The resonator is coupled to a transmission line through a coupling capacitor $C_{\rm g}$. The transmission line has a characteristic impedance $Z_{\rm tr} \approx 50\ \Omega$ and mean thermal occupation $N_{\rm tr}$ of photons at the frequency $\omega_{\rm r}$ corresponding to a temperature $T_{\rm tr}$. The coupling strength $\gamma_{\rm tr}$ of the resonator to the transmission line given by Eq.~\eqref{eq:gamma} was derived in Sec.~\ref{sec.tun}. With these results, we consider the energy exchange at the two interfaces of the resonator. The net power flowing from the SINIS junction to the resonator is $P_{\rm T}=\hbar \omega_{\rm r} \gamma_{\rm T} (N_{\rm T}-N_{\rm r})$ and the corresponding power from the resonator to the transmission line equals $P_{\rm tr}=\hbar \omega_{\rm r} \gamma_{\rm tr} (N_{\rm r}-N_{\rm tr})$. In thermal equilibrium and in the absence of other dissipation channels for the resonator, the powers balance each other, $P_{\rm tr}=P_{\rm T}$, the fact exploited in the experiments for probing quantum-circuit refrigeration and heating~\cite{Tan16, Masuda16}.

Expressing in the power $P_{\rm T}=\hbar \omega_{\rm r} \gamma_{\rm T} (N_{\rm T}-N_{\rm r})$ the resonator mean occupation $N_{\rm r}$ by means of Eq.~\eqref{eq:Nr}, we obtain
\begin{equation}
  P_{\rm T}=\hbar \omega_{\rm r}\frac{\gamma_{\rm T} \gamma_{\rm tr}}{\gamma_{\rm T}+\gamma_{\rm tr}} (N_{\rm T}-N_{\rm tr})\label{eq:PT}.
\end{equation}
If the net output power $P_{\rm T}=P_{\rm tr}$ is positive, the resonator is heated by the voltage biased SINIS junction above the constant reference value $N_{\rm tr}$, and vice versa for negative power. Above the gap, the temperature of the thermal reservoir increases linearly with $V$ and the coupling strength saturates to $\bar{\gamma}_{\rm T}$ [Eq.~\eqref{eq:above}], implying roughly linearly increasing output power as a function of the bias voltage beyond the gap. However, increased bias voltage also heats up the electrons in the normal metal,~\cite{Giazotto06} which leads to an additional weak reduction of the emission power.

To test the theory developed in this paper, we compute the net output power $P_{\rm T}$ of Eq.~\eqref{eq:PT} based on the thermal reservoir formulation of $\gamma_{\rm T}$ and $N_{\rm T}$ in Eq.~\eqref{eq:gammaN} using the parameters of the recent experimental work of Ref.~\onlinecite{Masuda16}. In Fig.~\ref{fig:MWS}, we achieve an excellent quantitative agreement between our theory and the experiments without fitting parameters, which provides a rigorous verification of the developed theory.

\subsection{Comparison with the $\mathbf{P(E)}$-theory}
\begin{figure}
  \centering
  \includegraphics[width=0.95\linewidth]{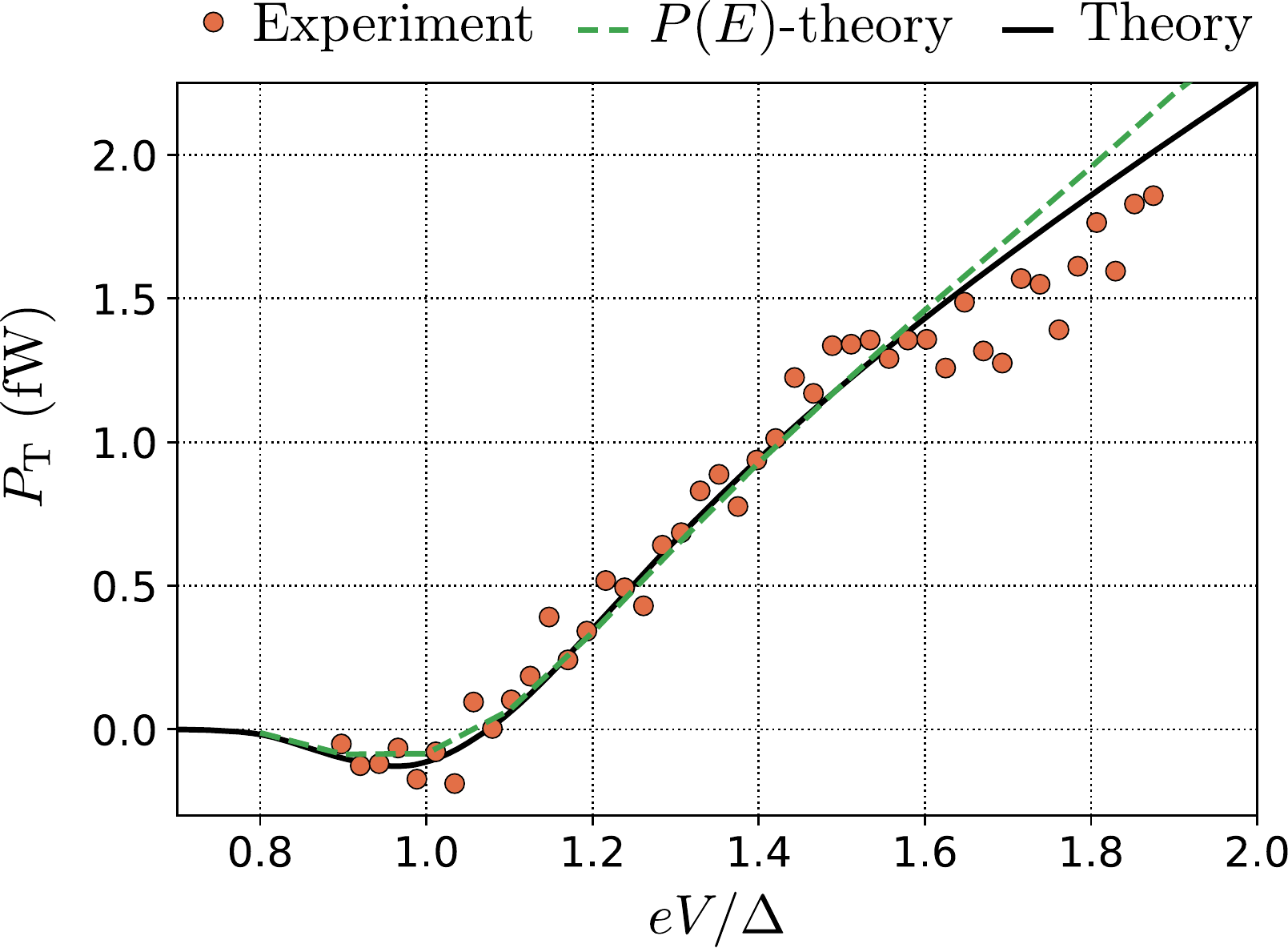}
  \caption{\label{fig:MWS}Net output power $P_{\rm T}$ as a function of the single-junction bias voltage. Comparison of Eq.~\eqref{eq:PT} providing the results of our transition rate theory (solid line) with the measured values (filled circles) and results of the $P(E)$-theory (dashed line) from Ref.~\onlinecite{Masuda16}. The parameters in the models correspond to the experiments: $\Delta=220$~$\mu$eV, $\gamma_{\rm D}=4\times 10^{-4}$, $R_{\rm T}=12.5$~k$\Omega$, $C_{\rm c}=840$~fF, $C_{\Sigma m}=12$~fF, $\omega_{\rm r}/2\pi = 4.55$~GHz, $Z_{\rm r}=33.9\ \Omega$, $C_{\rm g}=72.0$~fF, $Z_{\rm tr}=53.0\ \Omega$, and $T_{\rm tr}=180$~mK. For the electron temperature, we use the bias-voltage-dependent values $T_{\nmd}(V)$ experimentally measured in Ref.~\onlinecite{Masuda16} using an additional SINIS thermometer.}
\end{figure}

Previously, the $P(E)$-theory~\cite{IngoldNazarov05} has been used to obtain the tunneling-induced transition rates in a very good agreement with the experimental data both in the cooling~\cite{Tan16} and heating~\cite{Masuda16} regimes of the biased SINIS junction. In the $P(E)$-theory one considers a junction embedded in an electromagnetic environment formed by the surrounding electric circuitry. In addition to the occupations and densities of states at either side of a junction, the probability of a tunneling event depends on the ability of the environment to absorb the excess energy or to supply the remaining energy of a tunneling quasiparticle. This ability is described by the so-called $P(E)$-function, which is typically calculated for the surrounding electric circuitry in a thermal equilibrium (see Ref.~\onlinecite{Souquet14} for a nonequilibrium generalization).

The transition rate theory developed here agrees with the results of the $P(E)$-theory in the regime considered in Fig.~\ref{fig:MWS}. In Ref.~\onlinecite{Tan16}, the $P(E)$-theory was applied to derive the transition rates between the two lowest levels of a resonator in the case of a galvanic contact ($\alpha=1$) between the resonator and the normal-metal island. These results also agree with our results derived in Sec.~\ref{sec:Tunnel}. However, we note that the transition rate theory is considerably more general than $P(E)$-theory. It is able to capture conveniently the fine details of the physical circuit such as non-linearities of the electromagnetic environment of the tunnel junction. Furthermore, it is not restricted to surrounding electric circuits in thermal equilibrium but is directly applicable to non-equilibrium circuits including, \textit{e.g.}, driven superconducting qubits.

\section{Outlook and conclusions} \label{sec:conc}
We have developed a first-principles theory of quantum-circuit refrigeration by photon-assisted tunneling for a superconducting resonator. For weak interaction parameter $\rho$ characterizing the coupling strength of the tunneling transitions to the resonator, it leads to an intuitive thermal reservoir model of tunneling-induced transition rates of the resonator. This model allows us to condense the essential physics into the familiar concepts of the reservoir temperature and coupling strength to the resonator. In addition, we derived accurate analytical approximations for the temperature and the coupling strength in the experimentally relevant parameter regions. In the future, the thermal reservoir model can be straightforwardly extended to an input-output formulation~\cite{GardinerCollet85,Clerk10} describing, \textit{e.g.}, reflection of coherent microwaves from a refrigerated resonator.

Although we focus here on the effect of the quantum-circuit refrigerator on a linear resonator, the method of deriving the resulting transition rates is independent of the quantum circuit in question. In fact, the rates are products of an integral over Fermi functions and the matrix element of the charge shift operator between the eigenstates of the quantum circuit. A different quantum circuit simply yields different matrix elements. Thus our results can likely be generalized to the full spectrum of superconducting quantum devices. For example, the superconducting transmon qubit~\cite{Koch07, Paik11, Rigetti12, Barends2013} seems an ideal device to be cooled by the quantum-circuit refrigerator due to its efficient transverse coupling to charge and simultaneous exponentially suppressed sensitivity of the eigenenergies to charge offsets.

\begin{acknowledgements}
We acknowledge useful discussions with J.~Govenius, J.~Goetz, J.~Tuorila, and M.~Partanen. This research was financially supported by European Research Council under Grant No.~681311 (QUESS), by Academy of Finland under Grant Nos.~251748, 265675, 276528, 284621, 305237, 305306, 308161, and 312300, and by Alfred Kordelin Foundation.
\end{acknowledgements}
\appendix

\section{Island charge dynamics}\label{app:chargethermo}
 The master equation~\eqref{ME} shows that the charge state $\ket{q}$ of the normal-metal island is driven by both, elastic and inelastic transitions in the voltage-biased SINIS junction. Typically, the elastic transitions have a much higher rate than the inelastic ones (see Fig.~\ref{fig:gammaN}), originating from the fact that the matrix elements for elastic events scale as $M^2_{mm} \propto 1-\rho$ whereas the scaling for inelastic transitions is $M^2_{m m^\prime} \propto \rho^{|\ell|}$ with $\ell=m-m^\prime\geq 1$.

In this case, the master equation~\eqref{ME} rapidly leads to a steady-state distribution for the island charge by means of frequent elastic tunneling transitions. The resonator states equilibrate through slower inelastic processes or through transitions induced by coupling to the transmission line which is independent of the charge state. Thus we first compute the steady-state charge distribution $p_q$ determined by the elastic tunneling rates
\begin{equation}
\Gamma^\pm_{q,m,m}(V)=M^2_{mm}\frac{R_{\rm K}}{R_{\rm T}} \sum_{\tau=\pm 1} \normrateright\left ( \tau eV-E^\pm_q \right) \label{eq:Gamma0},
\end{equation}
and then compute the inelastic rates averaged over this distribution as in Eq.~\eqref{Gammaud}.
\begin{figure}
  \centering
  \includegraphics[width=0.95\linewidth]{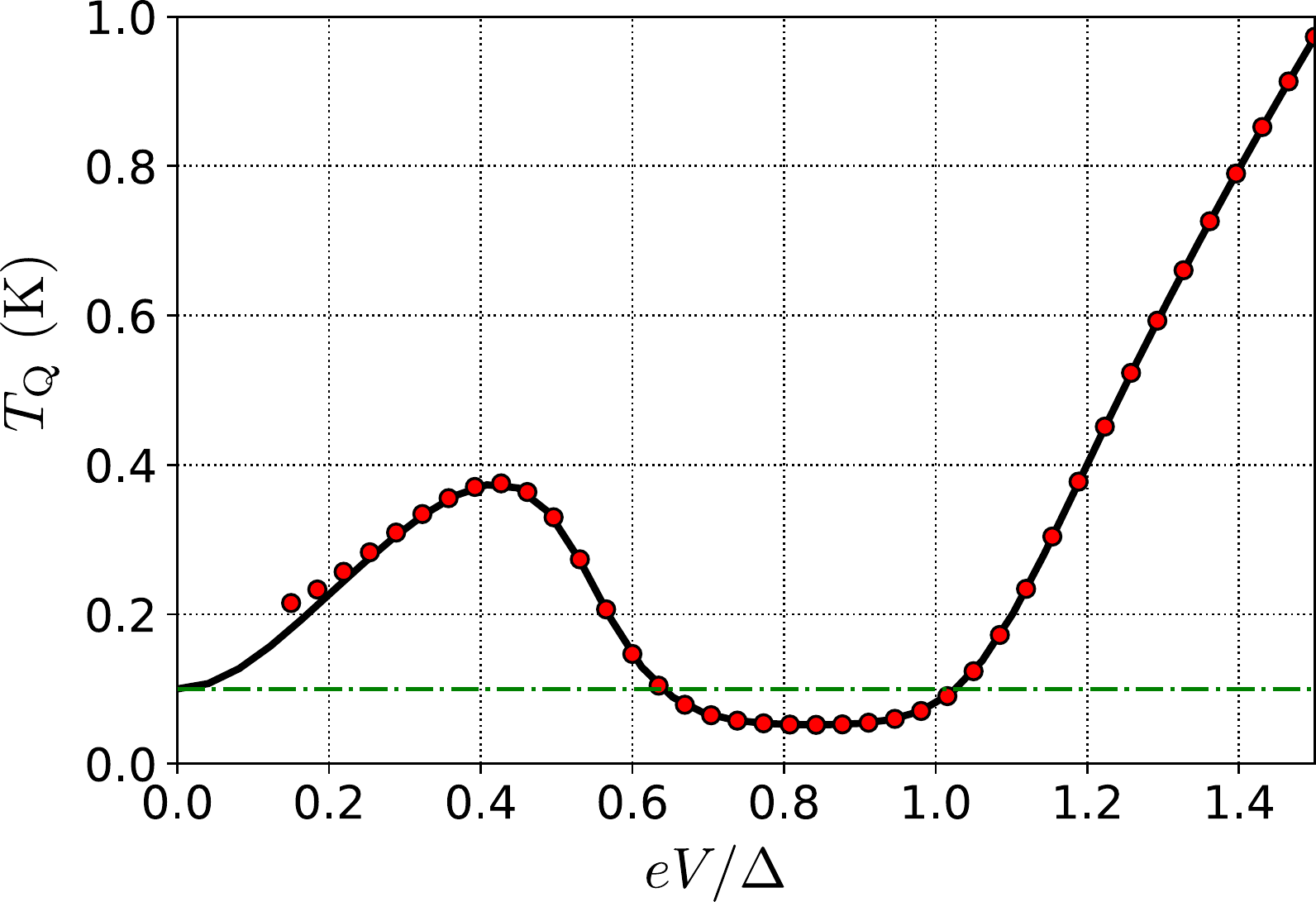}
  \caption{\label{fig:islandT}Effective temperature $T_{\rm Q}$ of the charge distribution of the normal-metal island as a function of the single-junction bias voltage. The parameters equal to those of Fig.~\ref{fig:gammaN}. The solid black line shows the numerical result based on Eqs.~\eqref{pqmsteady} and \eqref{eq.lambdaP}. The filled circles show the corresponding result based on Eqs.~\eqref{pqmsteady_expand}-\eqref{eq:gv} valid at $eV\gg E_{\nmd},k_{\rm B}T_{\nmd}$. In the deep-subgap regime, the temperature of the charge distribution $T_{\rm Q}$ is close to the electron temperature $T_{\nmd}=100$~mK depicted by a green dash-dotted line.}
\end{figure}

Note that in Eq.~\eqref{eq:Gamma0}, a change in the resonator state results in a simple scaling of the rates which affects only the speed of equilibration, not the actual steady-state distribution $p_q$. Thus for any resonator state indexed by $m$, the distribution can be accurately computed using the master equation
\begin{align}
\dot p_{q}(t)=\Big[\Gamma^+_{q\!-\!1, m, m}\,p_{q\!-\!1}(t)+\Gamma^-_{q\!+\!1, m, m}\,p_{q\!+\!1}(t)& \notag \\
-\left(\Gamma^+_{q, m, m}+\Gamma^-_{q, m, m}\right) & p_{q}(t)\Big].
\end{align}
This results in a steady-state distribution
\begin{equation}
  p_{q}=\frac{1}{Z} \exp\left[ \sum_{q^\prime=0}^{q-1} \ln\left(\frac{\Gamma^+_{q^\prime, m, m}}{\Gamma^-_{q^\prime \!+\!1, m, m}}\right)\right] \label{pqmsteady},
\end{equation}
where $q>0$ and $Z$ is a normalization factor. For $q<0$, we employ the symmetry $p_{-q}=p_q$. From Eq.~\eqref{eq:Gamma0} and the identity $E^-_{q+1}=-E^+_q$ we obtain
\begin{equation}
\frac{\Gamma^+_{q,m,m}}{\Gamma^-_{q\!+\!1,m,m}}
=\frac{\sum_{\tau=\pm 1} \normrateright\left ( \tau eV-E^+_q \right)}{\sum_{\tau=\pm 1} \normrateright\left ( \tau eV+E^+_q \right)}. \label{eq.lambdaP}
\end{equation}
In Fig.~\ref{fig:islandT}, we show the effective temperature of the charge distribution obtained from Eqs.~\eqref{pqmsteady} and~\eqref{eq.lambdaP} as a function of the single-junction bias voltage $V$ using experimentally relevant parameter values. The effective temperature $T_{\rm Q}$ is defined by a least-square fit of the thermal distribution $\widetilde{p}_{\rm q}=\exp[-E_{\nmd}q^2/(k_{\rm B}\widetilde{T}_{\rm Q})]/Z$ to the distribution $p_q$ of Eqs.~\eqref{pqmsteady} and \eqref{eq.lambdaP}.

Equation~\eqref{eq.lambdaP} shows that the steady state is determined by a ratio of  tunneling processes for energies differing by $2E^+_{\rm q}=2E_{\nmd}(1+2q)$. In the deep subgap regime, where $eV\ll k_{\rm B} T_{\nmd}$ and $eV\lesssim E_{\nmd}$, we may use the results of Sec.~\ref{sec:deepgap} and arrive at $T_{\rm Q}^{\rm dg}=T_{\nmd}$. At higher voltages, we utilize the fact that $E_{\nmd} \ll \Delta, k_{\rm B} T_{\nmd}, eV$ and expand Eq.~\eqref{eq.lambdaP} to the first order in $E^+_{q}$. Consequently, Eq.~\eqref{pqmsteady} results in a charge distribution of the thermal form
\begin{equation}
  p_{q}=\frac{1}{Z} \exp\left[-\frac{E_{\nmd} q^2}{k_{\rm B}T_{\rm Q}(V)} \right], \label{pqmsteady_expand}
\end{equation}
where $E_{\nmd} q^2$ corresponds to the energy of the charge state, $Z=\sum_q \exp[-E_{\nmd} q^2/(k_{\rm B}T_{\rm Q})]$ to the partition function, and $T_{\rm Q}(V)$ to the temperature of the distribution. The effective temperature assumes the form
\begin{equation}
 T_{\rm Q}(V)=\frac{1}{2k_{\rm B}}\left[\frac{g'(V)}{g(V)}-\frac{1}{k_{\rm B}T_{\nmd}\sinh\left(\frac{eV}{k_{\rm B}T_{\nmd}}\right)}\right]^{-1} \label{eq:bq},
\end{equation}
where the function $g(V)$ characterizes the difference of the Fermi functions weighted with the density of  states of the superconductor as
\begin{subequations} \label{eq:gv}
\begin{align}
  g(V) & =\int_{-\infty}^\infty \D{\varepsilon}\: n_{\rm S}(\varepsilon)\left[f(\varepsilon-eV)-f(\varepsilon)\right],\\
  g'(V) & =\frac{1}{4k_{\rm B}T_{\nmd}} \int_{-\infty}^\infty \D{\varepsilon}\ \textrm{sech}^2 \left (\frac{\varepsilon- eV}{2k_{\rm B}T_{\nmd}}\right) n_{\rm S}(\varepsilon).
\end{align}
\end{subequations}
Figure~\ref{fig:islandT} indicates that the first-order result of Eqs.~(\ref{pqmsteady_expand}) and~(\ref{eq:bq}) gives in general a very accurate approximation of the full numerical result of Eqs.~(\ref{pqmsteady}) and~(\ref{eq.lambdaP}).

\end{document}